\def\ls{\lower4pt\hbox{${\buildrel < \over \sim}$}}
\def\gs{\lower4pt\hbox{${\buildrel > \over \sim}$}}

\def\ox{\overline x}
\def\oxp{\overline x'}
\def\dtaccf{\overline{\Delta t}_{\rm acc, f}}

\def\betaf{\overline\beta_f}
\def\betar{\overline\beta_r}
\def\gup{\gamma_{\rm up}}
\def\gc{\gamma_c}
\def\gmin{\gamma_{\rm min}}
\def\gonef{\gamma_{1,f}}

\def\gtwof{\gamma_{2,f}}

\def\otx{\overline t_x}

\def\nuf{\nu_{\rm 0,f}}

\def\xcaplus{\overline x'_{\rm c, A+}}

\documentclass[12pt, preprint]{aastex}

\slugcomment{Submitted to {\it The Astrophysical Journal}}

\shorttitle{Internal Shocks in Blazars}
\shortauthors{M. B\"ottcher \& C. D. Dermer}

\begin{document}

\title{Timing Signatures of the Internal-Shock Model for Blazars}

\author{M. B\"ottcher\altaffilmark{1} and C. D. Dermer\altaffilmark{2}}

\altaffiltext{1}{Astrophysical Institute, Department of Physics and Astronomy, \\
Clippinger 339, Ohio University, Athens, OH 45701, USA}
\altaffiltext{2}{Naval Research Laboratory, Code 7653, Washington, D.C. 20375}

\begin{abstract}
We investigate the spectral and timing signatures of the internal-shock 
model for blazars. For this purpose, we develop a semi-analytical model 
for the time-dependent radiative output from internal shocks arising 
from colliding relativistic shells in a blazar jet. The emission 
through synchrotron and synchrotron-self Compton (SSC) radiation as 
well as Comptonization of an isotropic external radiation field 
are taken into account. We evaluate the discrete correlation 
function (DCF) of the model light curves in order to evaluate 
features of photon-energy dependent time lags and the quality 
of the correlation, represented by the peak value of the DCF. 
The almost completely analytic nature of our approach allows us to
study in detail the influence of various model parameters on the
resulting spectral and timing features. This paper focuses on a
range of parameters in which the $\gamma$-ray production is dominated
by Comptonization of external radiation, most likely appropriate for
$\gamma$-ray bright flat-spectrum radio quasars (FSRQs) or low-frequency
peaked BL Lac objects (LBLs). In most cases relevant for FSRQs and
LBLs, the variability of the optical emission is highly correlated
with the X-ray and high-energy (HE: $100$~MeV) $\gamma$-ray emission.
Our baseline model predicts a lead of the optical variability with
respect to the higher-energy bands by 1 -- 2 hours and of the
HE $\gamma$-rays before the X-rays by about 1 hour.
We show that variations of certain parameters may lead to changing 
signs of inter-band time lags, potentially explaining the lack of 
persistent trends of time lags in most blazars. 
\end{abstract}


\section{Introduction}

Blazars, a class of active galactic nuclei (AGNs) comprised of
Flat-Spectrum Radio Quasars (FSRQs) and BL~Lac objects, exhibit 
some of the most violent high-energy phenomena observed in AGNs 
to date. Their spectral energy distributions (SEDs) are characterized 
by non-thermal continuum spectra with a broad low-frequency component 
in the radio -- UV or X-ray frequency range and a high-frequency 
component from X-rays to $\gamma$-rays. They show rapid variability
across the electromagnetic spectrum. In extreme cases, the very-high-energy
(VHE) $\gamma$-ray emission of blazars has been observed to vary on
time scales of just a few minutes \citep{albert07,aharonian07}.

The flux variability of blazars is often accompanied by spectral
changes. Typically, the flux is most rapidly variable at the high-frequency 
ends of the two broad spectral components of the blazar SED. In the case 
of quasars, this refers to the optical (B-band) to UV and MeV to GeV $\gamma$-ray 
bands, while in the case of high-frequency peaked BL Lac objects (HBLs)
it is the X-ray and VHE $\gamma$-ray regimes where the variability is
the most extreme. In a few HBLs, the X-ray spectral variability could 
occasionally be characterized by spectral hysteresis, i.e., a loop 
track of the blazar's state in a hardness-intensity diagram 
\citep[e.g.][]{takahashi96,kataoka00,fossati00,zhang02}, although
even within the same object this feature tends not to be persistent
over multiple observations. Also in other types of blazars, hints 
of time lags between different observing bands are occasionally 
found in individual observing campaigns \citep[e.g.,][]{boettcheral07,horan09}, 
but the search for time-lag patterns persisting throughout multiple 
years has so far remained unsuccessful \citep[see, e.g.][for a 
systematic search for time lags between optical, X-ray and $\gamma$-ray 
emission in the quasar 3C279]{hartman01}.

In the framework of relativistic jet models, the low-frequency (radio
-- optical/UV) emission from blazars is interpreted as synchrotron
emission from nonthermal electrons in a relativistic jet. The
high-frequency (X-ray -- $\gamma$-ray) emission could either be
produced via Compton upscattering of low frequency radiation by the
same electrons responsible for the synchrotron emission \citep[leptonic
jet models; for a recent review see, e.g.,][]{boettcher07}, or 
due to hadronic processes initiated by relativistic protons 
co-accelerated with the electrons \citep[hadronic models, for 
a recent discussion see, e.g.,][]{muecke01,muecke03}. Leptonic 
models have been considered in a time-dependent manner with the 
aim of reproducing simultaneously the SEDs and light curve features 
of blazars \citep[see, e.g.,][]{kirk98,gm98,cg99,kataoka00,kusunose00,li00,bc02,jb07}.
The time-dependent analysis of homogeneous single-zone leptonic models
showed that spectral hysteresis patterns can be reproduced in a scenario 
of gradual particle acceleration and subsequent radiative cooling, and 
that the presence and direction of hysteresis patterns depends on the 
relative values of the time scales for particle acceleration, escape, 
and radiative cooling \citep{der98,cg99}. 

Homogeneous leptonic jet models have met with remarkable success
explaining the SEDs and correlated variability in many blazars.
However, several recent observational results have seriously 
challenged homogeneous models
and have motivated the consideration 
of inhomogeneous jet models. These observations include the 
uncorrelated variability between
X-rays and $\gamma$-rays 
in the HBLs 1ES1959+650 \citep{kraw04}
and Mrk~421 \citep{blaz05},
and the uncorrelated optical and TeV emissions in PKS~2155-304,
while X-rays and TeV $\gamma$-rays were well correlated \citep{costamante08}.
A particularly well motivated inhomogeneous blazar model is the
internal shock model \citep[e.g.][]{spada01,sokolov04,mimica04,graff08}. 
In this model, the central engine is intermittently ejecting shells 
of relativistic plasma at varying speeds, which subsequently collide. 
Such models have had remarkable success in explaining SEDs and time
lag features of generic blazars and deserve further study.

The realistic treatment of radiation transfer in an internal-shock
model for a blazar requires the time-dependent evaluation of retarded
radiation fields originating from all parts of the shocked regions 
of the jet. The model system is therefore highly non-linear and can
generally only be solved using numerical simulations 
\citep[e.g.,][]{sokolov04,mimica04,graff08,joshi09}. As the current 
detailed internal-shock models employ either full expressions or 
accurate approximations to the full emissivities of synchrotron 
and Compton emission, 
a complete simulation of the time-dependent 
spectra and light curves is
time-consuming and does therefore 
generally not allow to efficiently explore a large parameter 
space. General patterns of the SED, light curves and expected 
time lags between different wavelength bands have been demonstrated 
for very specific, but observationally very poorly constrained, 
sets of parameters. 

For this reason, we here develop a simplified internal-shock model in 
which the time-dependent synchrotron and external-Compton (EC) spectra 
are calculated completely analytically, and the SSC emission is 
reduced to a two-dimensional integral to be performed numerically.
This approach allows us to calculate time-dependent snapshot spectra
and light curves within a few minutes, and scan a large parameter space
for the resulting spectral and timing features.

Observational data of blazars often have limited, incomplete, and
irregular time sampling, which complicates the evaluation of inter-band
time lags and cross-correlations. The routinely used analysis method
designed to overcome these problems is the discrete correlation
function \citep[DCF, see][]{ek88}. Therefore, in order to produce
results directly comparable to observations, we subject our simulated
light curves to the same DCF technique, and evaluate predicted inter-band
time lags and the quality of the correlations, as represented by the peak
values of the DCF. 

We describe the general outline of our model in \S \ref{modelsetup}.
The dynamics of particle acceleration and cooling, and the resulting
space- and time-dependent particle distributions, will be derived in
\S \ref{particles}. In \S \ref{radiation} we describe our evaluation
of the time-dependent radiative output from the internal-shock model.
We present and discuss the results of a general parameter study in 
\S \ref{results}, and conclude in \S \ref{summary}. The appendices
contain some details of the rather cumbersome analytical integrations
required to evaluate the radiative output.

\section{\label{modelsetup}Model Setup and Shell Dynamics}

\begin{figure}[ht]
\plotone{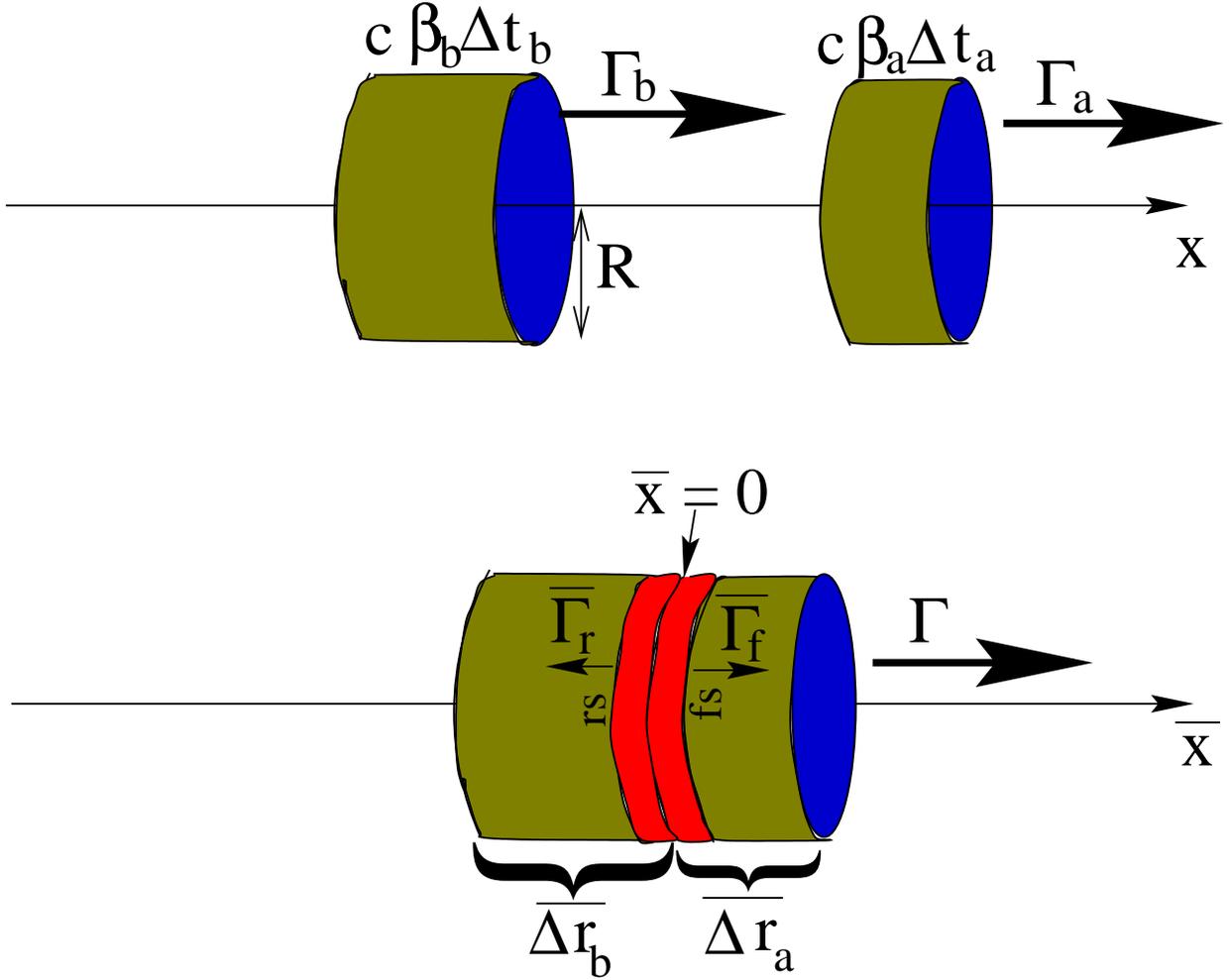}
\caption{Schematic of the colliding-shell geometry.}
\label{shells}
\end{figure}

We follow the collision of two relativistically moving shells (labeled
a and b) in a blazar jet, powered by an intermittent central source. 
The basic geometry is illustrated in Fig. \ref{shells}. \citep[For a 
related treatment
of a relativistic shell interacting with a shell of 
material at rest, see ][]{der08} The two shells 
are being ejected from the central engine with Lorentz factors $\Gamma_{a,b}$ 
with $\Gamma_b > \Gamma_a \gg 1$ and associated normalized velocities $\beta_{a, b}$
(with $\beta = v/c$). In the rest-frame of the central engine, the ejection events 
of the two shells last for a time $\Delta t_{a,b}$. Consequently, the shells 
have widths (in the central-engine rest frame) of $\Delta r_{a,b} = c \beta_{a,b}
\Delta t_{a,b}$. This assumes that the shells do not spread along or
transverse to the direction of motion. The former effect can be important 
at $r_{a,b}\gtrsim \Gamma_{a,b}^2 c\Delta t_{a,b}$ \citep[e.g.,][]{mlr93}, 
but is neglected because of
the short duration of the collision during 
which the shell can be approximated as having a nearly constant
thickness. Sideways expansion of the jet can also be neglected because 
of the short duration of the collision \citep[this effect can be important 
for narrow decelerating jets; see e.g.,][]{sph99}. This latter assumption 
is also supported by observations of extragalactic jets
remaining well 
collimated out to kpc scales.
If this assumption is valid, the shell
dynamics will not depend on the time between the shell ejections.

Assuming that before the collision, the kinetic energy of the plasma
in the shells' rest frames is negligible compared to the bulk kinetic 
energy (``cold plasma''), the kinetic luminosity of the shells is given 
by $L_{a,b} = \Gamma_{a,b} \dot{M}_{a,b} c^2$. This yields particle
densities in the shells of
\begin{equation}
{n'}_{a,b} = {L_{a,b} \over \pi R^2 \, \beta_{a,b} \Gamma_{a,b}^2 
\, m_p c^3}
\label{densities}
\end{equation}
where $R$ is the cross-sectional radius of the jet (and hence the shells),
and the prime denotes quantities in the shell rest frame.

The collision between the two shells will lead to the formation of a forward 
shock moving into the slower shell $a$, and a reverse shock moving back into
shell $b$. In the following, an overline denotes quantities in the frame
of the shocked material behind the shock fronts. In this frame, the 
forward and reverse shocks move with Lorentz factors $\overline\Gamma_{f,r}$,
respectively, away from the contact discontinuity. Pressure equilibrium
across the contact discontinuity yields the condition 

\begin{equation}
n'_a \left( \overline\Gamma_f^2 - \overline\Gamma_f \right) =
n'_b \left( \overline\Gamma_r^2 - \overline\Gamma_r \right).
\label{pressure}
\end{equation}
Denoting the bulk Lorentz factor of the shocked material with respect
to the stationary (AGN) rest frame as $\Gamma$ (with velocity $\beta$c), 
the shell velocities can be transformed to the shocked fluid frame through
\begin{equation}
\overline\Gamma_{f,r} = \Gamma_{a,b} \Gamma \, (1 - \beta_{a,b} \, \beta).
\label{Gamma}
\end{equation}
Eqs. \ref{pressure} and \ref{Gamma} can be solved to find $\Gamma$, 
$\overline{\Gamma_f}$ and $\overline{\Gamma_r}$. The solution is
obtained numerically \citep[cf.][]{sp95,pm99,der08}.

The widths of the shells in the shocked-fluid frame are given by
\begin{equation}
\overline{\Delta r}_{a,b} = {\Gamma_{a,b} \, c \, \Delta t_{a,b}
\over \overline\Gamma_{f,r}}
\label{drshell}
\end{equation}
where we have set $\beta_{a,b} = 1$ since $\Gamma_{a,b} \gg 1$. 
The time it takes for the shocks to cross their respective shells, 
can then be calculated as
\begin{equation}
\overline{\Delta t}_{f,r} = {\Gamma_{a,b} \, \Delta t_{a,b} \over
\overline\Gamma_{f,r} \, \overline\beta_{f,r}}
\label{tshock}
\end{equation}
in the shocked-fluid frame.

\section{\label{particles}Relativistic Particle Dynamics}

Relativistic particles are entering the shocked-fluid region with
collective Lorentz factors $\overline\Gamma_{f,r}$, respectively.
Upon shock crossing, the plasma will be compressed by a compression
ratio $r$. For the results presented in \S \ref{results}, we have
used $r = 4$. We assume that a fraction $\epsilon_B$ of the energy 
density in the shocked plasma will be contained in the magnetic 
field behind the shock fronts, which yields
\begin{equation}
B_{f,r} = \sqrt{ 8 \pi r \, \epsilon_B \, \left( \overline\Gamma_{f,r}^2
- \overline\Gamma_{f,r} \right) n'_{a,b} \, m_p c^2}
\label{Bfield}
\end{equation}
for the magnetic field strength behind the shocks.

First- and second-order Fermi processes will accelerate particles at the shock fronts.
We characterize the resulting injection of relativistic particles into
the shocked-fluid frame through the parameter $\epsilon_e$ giving the fraction of the shocked
plasma kinetic energy in the shocked-fluid frame that is transferred into
relativistic electrons. Thus the injection power in relativistic electrons
at the shock fronts is given by
\begin{equation}
{d E_{f,r} \over dt} \Biggr\vert_{\rm rel. e} = \epsilon_e \, \pi R^2 
\, m_p c^3 \, n'_{a,b} \, \overline\Gamma_{f,r} \overline\beta_{f,r} 
\, \left( \overline\Gamma_{f,r} - 1 \right).
\label{dEdtshock}
\end{equation}
The injection of relativistic electrons is described as a power-law in 
electron energy, $E_e = \gamma m_e c^2$, 
\begin{equation}
{dn_e \over d\gamma \, dt} \equiv Q(\gamma) = Q_0 \, \gamma^{-q} 
\, H(\gamma; \gamma_1, \gamma_2)
\label{Qinj}
\end{equation}
where the triple-argument Heaviside function is defined as
$H(x; a, b) = 1$ if $a \le x \le b$ and 0 otherwise, and all
$\gamma$'s refer to the shocked-fluid frame. We parameterize the
width of the acceleration zone within the shocked-fluid region as
a multiple $\Delta_{\rm acc}$ of the Larmor radius of a proton
with $\gamma_p = \overline\Gamma_{f,r}$ in the
shocked-fluid frame,
\begin{equation}
\overline{\Delta r}_{\rm acc; f,r} = \Delta_{\rm acc} \,
{\overline\Gamma_{f,r} \, m_p c^2 \over e \, B_{f,r}} \sim 3 \times 10^{6} 
\, \Delta_{\rm acc} \, \left( {B_{f,r} \over {\rm G}} \right)^{-1} \, 
\overline\Gamma_{f,r} \; {\rm cm}.
\label{Dracc}
\end{equation}
Consequently, the particle injection at any given point in the
shocked-fluid region will be active for a time

\begin{equation}
\overline{\Delta t}_{\rm acc; f,r} = \Delta_{\rm acc} \,
{\overline\Gamma_{f,r} \, m_p c \over e \, B_{f,r} \, \overline\beta_{f,r}} 
\sim 10^{-4} \, \Delta_{\rm acc} \, \left( {B_{f,r} \over {\rm G}} 
\right)^{-1} \, \, \overline\Gamma_{f,r} \, \overline\beta_{f,r}^{-1} \; 
{\rm s}.
\label{Dtacc}
\end{equation}
Particle acceleration will hence be active in a volume 
$\overline V_{\rm acc; f,r} = \pi R^2 \, \overline{\Delta r}_{\rm acc; f,r}$.
This allows us to evaluate the normalization of the particle injection
function (\ref{Qinj}) as

\begin{equation}
Q_{0; f,r} = {\epsilon_e \, \pi R^2 \, m_p c \, n'_{a,b} \, 
\overline\Gamma_{f,r} \overline\beta_{f,r} \, \left( \overline\Gamma_{f,r}
 - 1 \right) \over \overline V_{\rm acc; f,r} \, m_e} \times
\cases{ {q - 2 \over \gamma_1^{2 - q} - \gamma_2^{2 - q}} & if $(q \ne 2)$ \cr\cr
        {1 \over \ln \left( \gamma_2 \over \gamma_1 \right)} & if $q = 2$.}
\label{Qnorm}
\end{equation}

The high-energy cutoff can be obtained by balancing the acceleration 
time scale of electrons with the synchrotron loss time scale. Writing 
the acceleration time scale as a factor $a_{\rm acc}$ times the electron 
gyration time scale, $t_{\rm acc} = 2 \pi \, a_{\rm acc} \, \gamma \, 
m_e c^2 / (e \, B)$, the maximum electron Lorentz factor will be given 
by

\begin{equation}
\gamma_{2; f,r} = \sqrt{3 \, e \over a_{\rm acc} \, \sigma_T \, B_{f,r}} 
\approx 4.6 \times 10^7 \, a_{\rm acc}^{-1/2} \, \left( {B_{f,r} \over
{\rm G}} \right)^{-1/2}
\label{gamma2}
\end{equation}

If a fraction $\zeta_e$ of electrons behind the shock fronts is
accelerated into the power-law distribution (\ref{Qinj}) and we
assume $\gamma_2 \gg \gamma_1$, then the low-energy cutoff of
the injection function (\ref{Qinj}) is given by
\begin{eqnarray}
&\gamma_{1; f,r} =
\cases{
{m_p \over m_e} \, {q - 2 \over q - 1} \, {\epsilon_e \over
\zeta_e} \, (\overline\Gamma_{f,r} - 1) & if $q > 2$ \cr\cr
\left( {m_p \over m_e} \, {2 - q \over q - 1} \, {\epsilon_e \over
\zeta_e} \, [\overline\Gamma_{f,r} - 1] \, \gamma_{2; f,r}^{q - 2}
\right)^{1 \over q - 1} & if $1 < q < 2$ \cr\cr
{m_p \over m_e} \, {\epsilon_e \over \zeta_e} \, 
(\overline\Gamma_{f,r} - 1)/\ln\left( {\gamma_{2; f,r} \over \gamma_{1; f,r}}
\right)& if $q = 2$\cr} \cr
\label{gamma1}
\end{eqnarray}
where the $q = 2$ case will be solved numerically.

The electrons injected according to Eq. \ref{Qinj} will subsequently
cool primarily due to radiative losses. Adiabatic losses become important 
after
the shocks have traversed the shell and the shocked fluid shell 
begins to expand. When this happens, the magnetic field and therefore 
the synchrotron emission rapidly decays \citep{der08}. We neglect 
adiabatic losses in this study because of the rapid radiative cooling 
of the electrons. A good approximation to the 
time-dependent shape of the electron distribution can be found if
all radiative losses can be described by a loss term

\begin{equation}
\dot\gamma = - \nu_0 \gamma^2
\label{gammadot}
\end{equation}
which holds for synchrotron emission in a constant magnetic field
as well as Compton scattering in the Thomson regime in a radiation
field of constant energy density. In that case,

\begin{equation}
\nu_0 = {4 \over 3} \, c \, \sigma_T \, {u \over m_e c^2}
\label{nu0}
\end{equation}
where $u$ is the sum of the magnetic-field and radiation field energy
densities in the shocked-fluid frame. If the system under consideration
is (a) SSC dominated and/or (b) Compton scattering at any given electron 
energy is predominantly happening in the Klein-Nishina regime, the 
evolution of the electron distribution can only be solved numerically. 
This is because in case (a), the dominant radiation field
for Compton cooling depends on the current (and recent) electron energy 
distribution and thus the cooling becomes non-linear \citep[see, e.g.,][for
an analytical treatment of non-linear radiative cooling]{rs09,schlick09}, 
and in case (b), 
the cooling curve flattens towards higher energies compared to the simple 
$\gamma^2$ dependence in Eq. \ref{gammadot}. Therefore, our analysis is
most directly applicable to the FSRQs and LBLs in which the $\gamma$-ray 
emission
is generally believed to be dominated by EC scattering, mostly 
in the
Thomson regime. 

If the conditions of a constant energy density $u$ and Compton scattering
in the Thomson regime are approximately fulfilled, then the time-dependent
electron distribution can be found analytically. For this purpose, we define
a spatial coordinate $\overline x$ which is defined in the shocked-fluid
frame as $\overline x = 0$ at the contact discontinuity, and is measured 
positive in the direction of the forward-shock propagation (i.e., into 
shell $a$; see Fig. \ref{shells}). At any given point $\overline x$, 
the time $\overline t_x$ elapsed since the onset of the acceleration 
(i.e., since the forward  or reverse shock has passed this point), is 
given by

\begin{equation}
\overline t_x = \overline t - { \vert \overline x \vert \over \overline 
\beta_{f,r} \, c}
\label{tx}
\end{equation}
where $\overline t$ denotes the reference time in the shocked-fluid frame.
At any given point, the acceleration will remain active for a time
$\overline{\Delta t}_{\rm acc; f,r}$. Then, the time- and space-dependent
relativistic, non-thermal electron distribution can be well approximated 
by

$$ \overline n (\gamma; \overline x, \overline t_x) = Q_0 \, 
H(\overline t_x) \, H(\gamma_{\rm up} - \gamma) \, \Biggl\lbrace
\min(\overline t_x, \overline{\Delta t}_{\rm acc; f,r}) \, \gamma^{-q}
\, H(\gamma; \gamma_1, \gamma_c) $$
\begin{equation}
+ {\min(\overline t_x, \overline{\Delta t}_{\rm acc; f,r}) \over \nu_0 \,
\overline t_x} \, \gamma^{-(1 + q)} \, H(\gamma - \gamma_1) \, H(\gamma; 
\gamma_c, \gamma_{\rm up})  \; + \; 
\overline{\Delta t}_{\rm acc; f,r} \, \gamma_1^{-q} \, \gamma^{-2}
H(\gamma; \gamma_{\rm min}, \gamma_1) \Biggr\rbrace
\label{ne}
\end{equation}
where the single-argument Heaviside function is defined as
$H(x) = 1$ if $x > 0$ and 0 otherwise, and the characteristic
electron energies are given by

\begin{eqnarray}
\gamma_{\rm up} &=& {1 \over \gamma_2^{-1} + \nu_0 \, \max\lbrace 0, (\overline t_x - 
\overline{\Delta t}_{\rm acc; f,r}) \rbrace} \cr
\gamma_c &=& {1 \over \nu_0 \overline t_x}, \cr
\gamma_{\rm min} &=& {1 \over \gamma_1^{-1} + \nu_0 \overline t_x}. \cr
\label{gammas}
\end{eqnarray}

\section{\label{radiation}Radiative Output}

The evaluation of the radiative output from the entire shock structure
will involve an integral of the emissivity $j_{\overline\epsilon}$
along the $\overline x$ direction:

\begin{equation}
\nu F_{\nu} (\epsilon, t_{\rm obs}) = {D^4 \, \pi R^2 \over d_L^2} \; 
\int\limits_{\overline x_{\rm min}}^{\overline x_{\rm max}}
\overline\epsilon \, j_{\overline\epsilon} (\overline x, \overline t_{\rm
x, em}) \; d \overline x
\label{nuFnuint}
\end{equation}
where $D = (\Gamma \, [1 - \beta \mu])^{-1}$ is the Doppler boosting factor
with $\mu = cos\theta$, the cosine of the viewing angle between the jet axis 
and the line of sight, and $\overline\epsilon = \epsilon \, (1 + z) / D$. 
The emission time $\overline t_{\rm x, em}$ has to be evaluated accounting 
for the light-travel time difference between different parts of the jet. 
The integration at a given observer's time 
$t_{\rm obs}$ has to be performed in a way that $dt_{\rm obs} = (1 + z) \, 
(dt - dx \, \mu / c)  = 0$, where $t$ and $x$ are measured in the 
stationary AGN frame. With $d\overline t = dt/\Gamma$ and $d\overline x
= \Gamma dx$, we find that an advancement in $d\overline x$ corresponds
to an advancement in emission time as $d\overline t = d\overline x \, \mu
/ (\Gamma^2 c)$. Consequently, the co-moving emission time elapsed since
the shock-crossing at any given point $\overline x$ at observed time 
corresponding to co-moving time $\overline t$ will be

\begin{equation}
\overline t_{\rm x, em} = \overline t - {\vert \overline x \vert \over
\overline\beta_{f,r} c} + {\overline x \, \mu \over \Gamma^2 c}.
\label{tem}
\end{equation}

The limits $\overline x_{\rm min,max}$ of the integration in Eq.
\ref{nuFnuint} will be given by constraints on the emission time
being $> 0$ (i.e., the shock front must have passed the point $\ox$), 
and the thickness of the shell:

\begin{eqnarray}
\overline x_{\rm min} &=& -\min\left( {\overline t \, \overline \beta_r \, c 
\over 1 + \mu \overline\beta_r / \Gamma^2} \; , \; \overline{\Delta r}_b
\right)  \cr\cr
\overline x_{\rm max} &=& \min\left( {\overline t \, \overline \beta_f \, 
c \over 1 - \mu \overline\beta_f / \Gamma^2} \; , \; \overline{\Delta r}_a 
\right)
\label{xlimits}
\end{eqnarray}

There will be the additional constraint of particles being available 
to contribute to the emission at a given energy $\overline\epsilon$, 
which depends on the individual emission mechanisms considered below. 

\subsection{\label{synchrotron}Synchrotron Emission}

For the purpose of our analytical treatment, we use a simple
$\delta$-function approximation for the synchrotron emissivity:

\begin{equation}
j_{\overline\epsilon, {\rm sy}} = {c \, \sigma_T \, B^2 \, \overline\epsilon
\over 48 \pi^2 \, b^2 \, \gamma_{\rm sy}} \; n_e (\gamma_{\rm sy})
\label{deltasy}
\end{equation}
where $b = B / B_{\rm crit}$ with $B_{\rm crit} = (m_e c^3)/(e \hbar)
\approx 4.4 \times 10^{13}$~G, and $\gamma_{\rm sy} = \sqrt{\overline\epsilon 
/ b}$. Inserting this into Eq. \ref{nuFnuint} yields

\begin{equation}
\nu F_{\nu}^{\rm sy} (\epsilon, t_{\rm obs}) = {D^{5/2} \over \sqrt{1 + z}}
\, {c \, \sigma_T \, B^2 \, R^2 \over 48 \, \pi \, d_L^2 \, b^{3/2}}
\, \epsilon^{3/2} \; \int\limits_{\overline x_{\rm min}}^{\overline 
x_{\rm max}} n_e \left(\sqrt{\epsilon \, (1 + z) \over b \, D} \, , \,
\overline t_{\rm x, em}\right) \; d \overline x
\label{nuFnusy}
\end{equation}
where we use Eq. \ref{ne} for the space- and time-dependent particle
distribution $n_e (\gamma_{\rm sy} , \overline t_{\rm x, em})$. The integral
in \ref{nuFnusy} can be solved fully analytically, and the solution is derived
in Appendix \ref{synchrotronapp}.

In this study, we focus on predictions for optical and higher-frequency emission. 
For this reason, we neglect synchrotron-self absorption (SSA) in our analysis. 
The emission region of our model system becomes optically thick to SSA at frequencies
well below the optical regime. In our model systems considered in \S \ref{results}, 
the SSA frequency (where $\tau_{\rm SSA} = 1$) is typically $\lesssim \nu_{\rm SSA} 
\sim 10^{13}$~Hz.

\subsection{\label{EC}External-Compton Emission}

For the purpose of an analytical treatment, we evaluate the Comptonization of
external radiation also with a simple $\delta$-function approximation for the
Thomson cross section. Furthermore, we assume that the external radiation field
is isotropic in the stationary AGN frame and characterize it as mono-energetic 
with frequency $\nu_{\rm ext}$, corresponding to a dimensionless photon energy 
in the co-moving frame, $\overline\epsilon_s = \Gamma \, h \nu_{\rm ext} / (m_e c^2)$. 
The radiation energy density $u_{\rm ext}$ in the stationary AGN frame will be boosted
into the shocked-fluid frame as $\overline u_{\rm ext} = \Gamma^2 u_{\rm ext}$.
The effect of the Klein-Nishina decline of the Compton cross section is approximated
as a hard cutoff in the scattered photon spectrum at $\epsilon_c = (D/[1 + z] \, 
\overline\epsilon_s)$. The beaming patterns of the intrinsically isotropic synchrotron
and the external-Compton emissions are slightly different because the external
radiation field is anisotropic in the co-moving frame \citep{dermer95}. However,
as long as the observer is located within the beaming cone at $\theta_{\rm obs}
\sim 1/\Gamma$, the difference is small and will be neglected in our simplified
semi-analytical treatment. With these approximations, the treatment of the external
Compton radiation is completely analogous to the one of synchrotron emission
(see Appendix \ref{synchrotronapp}), with the substitutions $B^2/(8 \pi) \to 
\overline u_{\rm rad}$ and $b \to \overline\epsilon_s$:

\begin{equation}
\nu F_{\nu}^{\rm EC} (\epsilon, t_{\rm obs}) = {D^{5/2} \over \sqrt{1 + z}}
\, {c \, \sigma_T \, \overline u_{\rm rad} \, R^2 \over 6 \, d_L^2 \, 
\overline\epsilon_s^{3/2}}
\, \epsilon^{3/2} \, H\left( {D \over (1 + z) \, \overline\epsilon_s} - 
\epsilon \right) \,  \int\limits_{\overline x_{\rm min}}^{\overline 
x_{\rm max}} n_e \left(\sqrt{\epsilon \, (1 + z) \over \overline\epsilon_s \, D} 
\, , \, \overline t_{\rm x, em}\right) \; d \overline x
\label{nuFnuec}
\end{equation}
with the solution to the integral given by the sum of the terms $I_{ir} + I_{if}$
($i = 1, 2, 3$) for $\gamma = \sqrt{(\epsilon \, (1 + z) / [\overline\epsilon_s
\, D])}$ derived in Appendix \ref{synchrotronapp}.

\subsection{\label{SSC}Synchrotron Self-Compton Emission}

For the evaluation of the synchrotron-self-Compton emissivity $j_{\overline\epsilon,
{\rm SSC}}$, to use in Eq. \ref{nuFnuint}, we adopt, again, a $\delta$-function
approximation for the Compton cross section,

\begin{equation}
{d\sigma \over d\epsilon_c \, d\Omega_c} \approx \sigma_T \, \delta (\epsilon_c
- \gamma^2 [1 - \beta\overline\mu_c] \epsilon_s) \, \delta(\Omega_c - \Omega_e) \;
H(1 - \gamma\epsilon_s [1 - \beta\overline\mu_c]).
\label{thomson}
\end{equation}
where $\overline\mu_c$ is the cosine of the collision angle between the scattering
electron and the incoming soft photon with energy $\epsilon_s$. In Eq. \ref{thomson}
the effect of the Klein-Nishina decline of the Compton cross section is incorporated 
as a hard cutoff at the transition from the Thomson to the Klein-Nishina regime. With 
these approximations, the Compton emissivity becomes

\begin{equation}
j_{\overline\epsilon, {\rm SSC}} (\overline x, \overline t_{\rm x, em})
\approx {c \, \sigma_T \, m_e c^2 \over 8 \pi} 
\overline\epsilon^{1/2} \int\limits_{4\pi} d\overline\Omega_s \, 
\int\limits_0^{1/(\overline\epsilon [1 - \overline\mu_c])} d\overline\epsilon_s
\, \sqrt{1 - \overline\mu_c} \, {\overline n_{\rm ph} (\overline\epsilon_s,
\overline\Omega_s, \overline x, \overline t_{\rm x, em}) \over 
\overline\epsilon_s^{1/2}} \, n_e (\gamma_c, \overline x, \overline t_{x, em}) 
\label{jssc1}
\end{equation}
with 

\begin{equation}
\gamma_c = \sqrt{ \overline\epsilon \over (1 - \beta\overline\mu_c) \, 
\overline\epsilon_s } 
\label{gammac}
\end{equation}

For relativistic electrons, $\gamma \gg 1$ we can neglect the factor $\beta$
in Eq. \ref{gammac} as long as $\overline\mu_c \ll \beta \sim 1$, so that it 
is an explicit solution. In order to use Eq. \ref{jssc1}, we need an expression 
for the synchrotron photon density $\overline n_{\rm ph} (\overline\epsilon_s, 
\overline\Omega_s)$, which is a convolution of the (retarded) contributions from 
all shocked parts of the jet. For simplicity, we assume that all synchrotron 
photons enter a given point along the jet either directly from the front 
(superscript `+') or from the back (superscript `-'). In those cases, the 
scattering angle $\overline\mu_c$ is determined through the angular $\delta$ 
distribution in Eq. \ref{thomson} as

\begin{equation}
\overline\mu_c^{\pm} = \mp \overline\mu = \mp {\mu - \beta_{\Gamma} \over 
1 - \beta_{\Gamma}\mu}
\label{mubar}
\end{equation}
where $\mu$ is the cosine of the observing angle in the observer's frame and
$\beta_{\Gamma} = \sqrt{1 - 1/\Gamma^2}$. We then write the synchrotron photon 
density as

\begin{equation}
\overline n_{\rm ph} (\overline\epsilon_s, \overline\Omega_s, \overline x,
\overline t_{\rm x, em}) \equiv
\overline n_{\rm ph}^+ (\overline\epsilon_s, \overline x, \overline t_{\rm x, em}) 
{\delta (\mu_c + \overline\mu) \over 2 \pi} 
+ \overline n_{\rm ph}^- (\overline\epsilon_s, \overline x, \overline t_{\rm x, em}) 
{\delta (\mu_c - \overline\mu) \over 2 \pi}
\label{nph1}
\end{equation}
where now the evaluation of $\overline n_{\rm ph}^{\pm} (\overline\epsilon_s)$
involves an integral over the retarded emission from the shocked parts of the 
jet in front of and behind the point under consideration, respectively. This 
is requires the evaluation of an integral

$$
\overline n_{\rm ph}^{\pm} (\overline x) = {1 \over 4 \pi \, c} \int\limits_{\overline 
x_{\rm s, min}^{\pm}}^{\overline x_{\rm s, max}^{\pm}} d\oxp \, \dot n_{\rm ph} 
(\oxp) \, \int\limits_0^R \, {r \, dr  \over ([\ox - \oxp]^2 + r^2)}
$$
\begin{equation}
= {1 \over 8 \pi \, c} 
\int\limits_{\overline 
x_{\rm s, min}^{\pm}}^{\overline x_{\rm s, max}^{\pm}} d\oxp \, \dot n_{\rm ph} (\oxp)
\ln\left( {(\ox - \oxp)^2 + R^2 \over (\ox - \oxp)^2} \right)
\label{nphintegral}
\end{equation}
The physically relevant case at hand here will correspond to thin slabs with
$R \gg \overline{\Delta r}_{\rm a,b}$, for which the integral simplifies to

\begin{equation}
\overline n_{\rm ph}^{\pm} (\overline x) = {1 \over 4 \pi \, c} \int\limits_{\overline 
x_{\rm s, min}^{\pm}}^{\overline x_{\rm s, max}^{\pm}} d\oxp \, \dot n_{\rm ph} (\oxp) 
\, \ln\left( {R \over \vert \ox - \oxp \vert } \right)
\label{nphlimit}
\end{equation}

The photon density distributions $\overline n_{\rm ph}^{\pm}$ are evaluated
fully analytically, as discussed in Appendix \ref{sscphotons}, while the
remaining two integrations over $\overline \epsilon_s$ and $\overline x$ 
will be done numerically. Those are the only numerical integrations needed
in our evaluation of the time-dependent emission spectra and light curves
from the internal shock model.

In our analysis, we neglect second-order SSC emission. This is justified by
(a) the small Thomson depth ($\sim 10^{-4}$) of our model systems considered
in the following section, and (b) the Klein-Nishina suppression of higher-order 
SSC emission.

\begin{figure}[ht]
\plotone{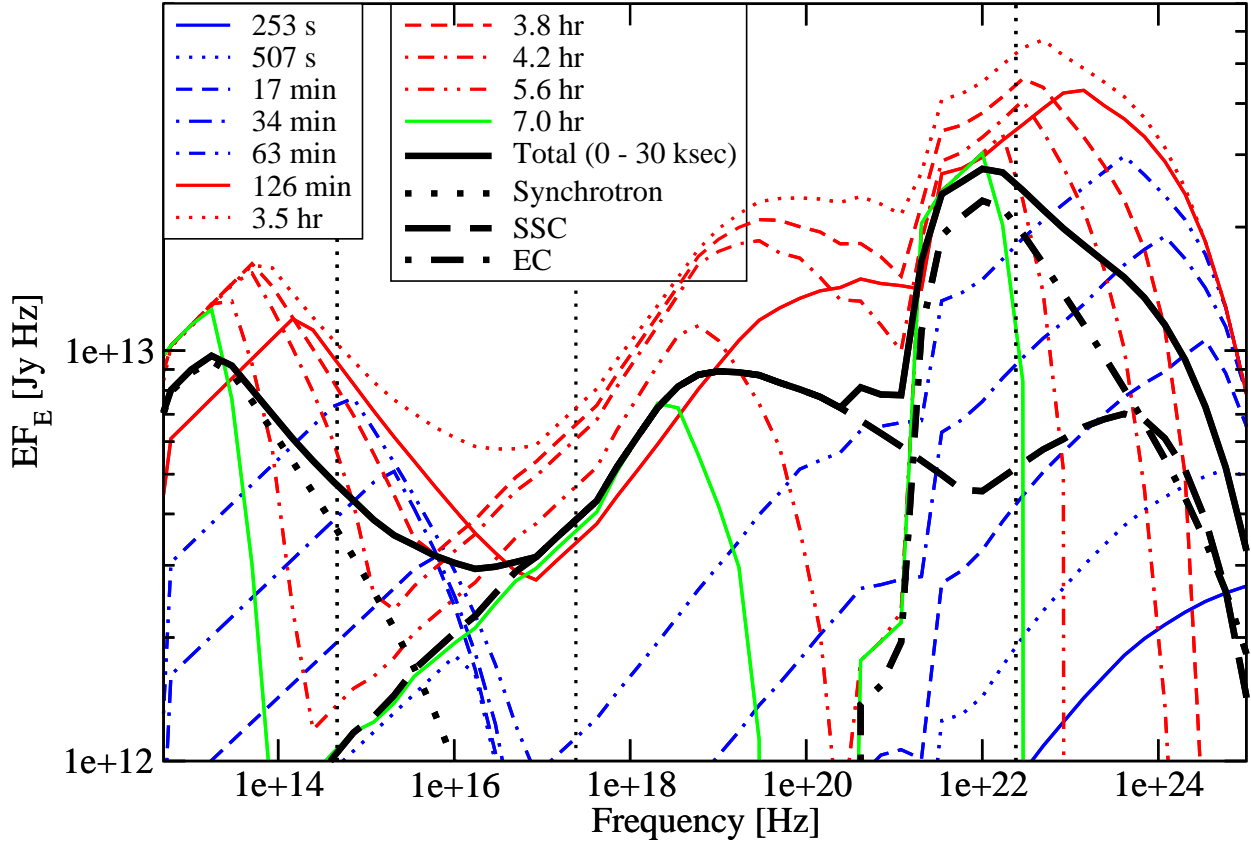}
\caption{Snap-shot SEDs for our baseline model with parameters listed
in Table \ref{baselinepars}. The heavy solid (red) curve shows the 
SED resulting from averaging over an integration time of 30~ksec,
representative of a typical exposure time of a deep X-ray observation 
of a blazar. Individual radiation components of the time-averaged
SED are shown as: dotted = synchrotron; long-dashed = SSC; dot-dashed
= EC. The dotted vertical lines indicate the frequencies (R-band, 1~keV,
1~MeV, 100~MeV) at which light curves have been extracted. }
\label{SEDs}
\end{figure}

\section{\label{results}Results}

We have applied the semi-analytical internal-shock model described in the
previous 
section, to parameter sets typical for FSRQs and LBLs. In these objects, the
$\gamma$-ray emission is generally believed to be dominated by Compton scattering
of external photons in the Thomson regime, in which case our approximate description 
of a constant electron cooling coefficient $\nu_0$ is valid. Table \ref{baselinepars} 
lists parameters which we adopt for a baseline model in this regime. The chosen model
parameters result in a bulk Lorentz factor of the shocked fluid of $\Gamma = 18.2$,
and a relative Lorentz factor between the shells of $\Gamma_{\rm rel} = 1.133$.
The observing angle $\theta_{\rm obs}$ has been chosen to coincide with the
angle for which $D = \Gamma$, and we take the $\epsilon_e$ and $\epsilon_B$ 
parameters equal in the forward and reverse
shocked regions.

\begin{deluxetable}{ccccc}
\tabletypesize{\scriptsize}
\tablecaption{Parameters of our baseline model.}
\tablewidth{0pt}
\tablehead{
\colhead{Parameter} & \colhead{Symbol} & \colhead{Value} }
\startdata
Lorentz factor of shell a	& $\Gamma_a$		& $15$ \\
Lorentz factor of shell b	& $\Gamma_b$		& $25$ \\
Kinetic power of shell a	& $L_a$			& $10^{49}$~erg~s$^{-1}$ \\
Kinetic power of shell b	& $L_b$			& $10^{49}$~erg~s$^{-1}$ \\
Duration of ejection of shell a	& $\Delta t_a$		& $2 \times 10^3$~s \\
Duration of ejection of shell b	& $\Delta t_b$		& $2 \times 10^3$~s \\
Observing angle 		& $\theta_{\rm obs}$	& $3.15^o$ \\
Shell radius 	 		& $R$			& $3 \times 10^{16}$~cm \\
Electron equipartition fraction	& $\epsilon_e$		& $0.1$ \\
B-field equipartition fraction	& $\epsilon_B$		& $10^{-3}$ \\
Electron acceleration fraction  & $\zeta_e$		& $0.01$ \\
Acceleration time scale par.	& $a_{\rm acc}$		& $10^6$ \\
Acceleration length parameter	& $\Delta_{\rm acc}$	& $10$ \\
Elect. injection spectral index & $q$			& $2.6$ \\
External rad. energy density 	& $u_{\rm ext}$	  	& $3 \times 10^{-4}$~erg~cm$^{-3}$ \\
External rad. peak frequency	& $\nu_{\rm ext}$	& $10^{14}$~Hz \\
Redshift			& $z$			& $0.5$ \\
\enddata
\label{baselinepars}
\end{deluxetable}

Fig. \ref{SEDs} shows the instantaneous broadband spectra from this baseline model. 
In order to compare the SEDs with observations typically requiring extended exposure 
times, we also evaluate the average SED over an integration time of 30~ksec. 
This time is representative of a deep blazar observation in X-rays, though 
shorter than typically required for Fermi to accumulate a signal with useful 
photon statistics from a bright $\gamma$-ray blazar. For Fermi, the minimum 
observation time from signal-dominated statistics for the detection of $5N_5$ 
photons at 1~GeV photon energy is $t_{obs} \gtrsim 25 N_5/(X_{1/5} \cdot 
8000{\rm~cm}^2 \cdot 10^{-10}f_{-10}$ erg cm$^{-2}$ s$^{-1}
\cdot 1.6\times 
10^{-3}$ ergs/GeV) $\approx 5\times 10^{4} N_5 / (X_{1/5} f_{-10})$~s, where 
$0.2 \, X_{1/5}$ is the fraction of time that Fermi will effectively observe 
any given target in scanning mode covering about $2.4$ sr \citep{atw09}.
For the bright flare parameters used in Fig. 2, the $\nu F_{\nu}$
flux at 1~GeV reaches values of $\approx 5 \times 10^{13}$~Jy~Hz or
$f_{-10} \sim 5$, so that variability as short as $\approx 3$~hours
could be detected. For comparison, the brightest blazar flare yet detected
with Fermi was from 3C~454.3, which reached flux levels $\gtrsim 2 \times
10^{-5}$~ph ($> 100$~MeV)~cm$^{-2}$~s$^{-1}$ \citep{et09}, somewhat dimmer
than the example considered here.

The time-averaged SED is illustrated by the heavy solid curve in Fig. 
\ref{SEDs}. It displays a
synchrotron peak in the infrared, 
at $\sim 2 \times 10^{13}$~Hz ($\lambda \sim 
15 \mu$m), as well as an EC dominated $\gamma$-ray peak at $\sim 10^{22}$~Hz
($\sim 40$~MeV), with the $\gamma$-ray flux dominating over the synchrotron 
peak flux. It also displays a very hard, SSC-dominated X-ray spectrum ($\alpha_X
< 1$). These SED properties are representative of blazars of the FSRQ
subclass, e.g., PKS~1510-089 \citep{ammando09} or 3C~454.3 \citep{abd09b}. 
We note that due to the hard low-energy cutoff introduced by our delta
function approximation to the Compton emissivity, the emission in the
hard X-ray -- soft $\gamma$-ray ($\sim 100$~keV -- a few~MeV) region may 
be underproduced in our simulations.

\begin{figure}[ht]
\plotone{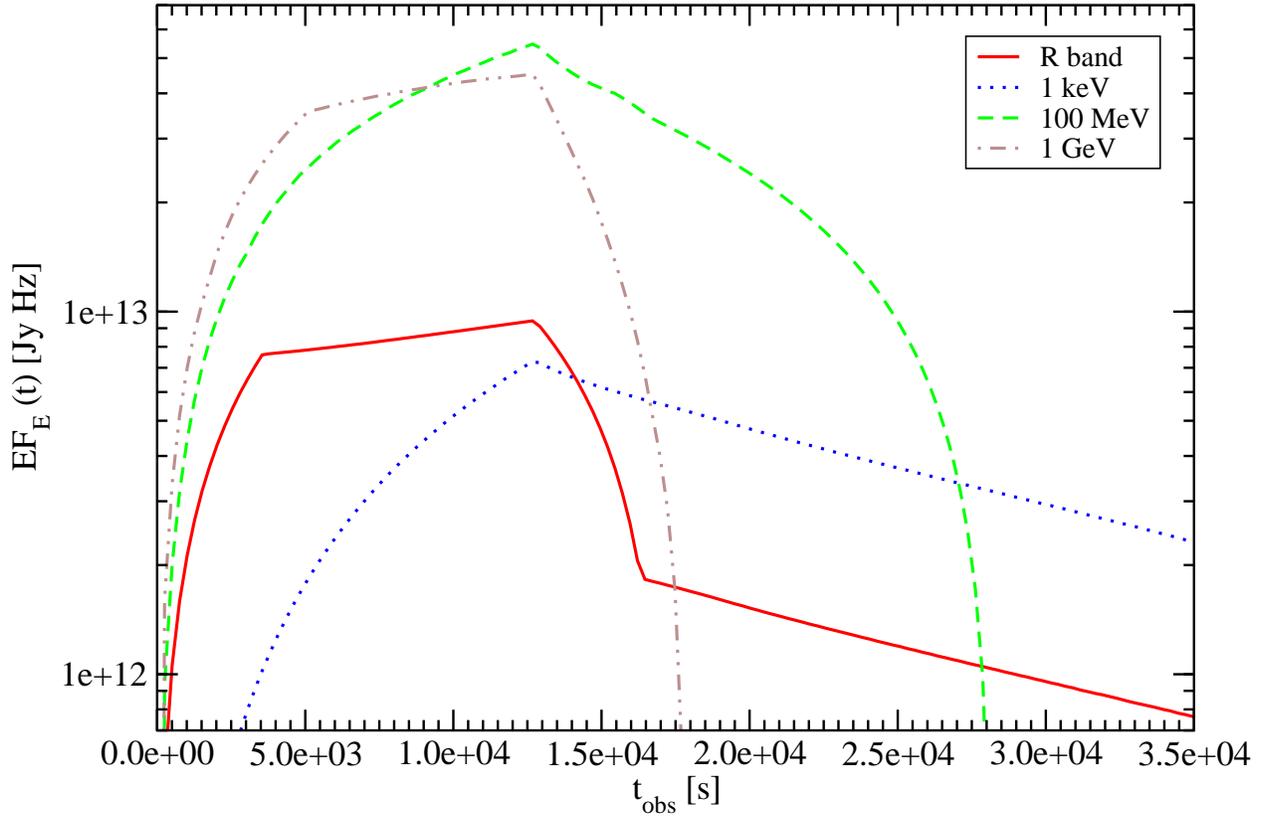}
\caption{Light curves at various frequencies/energies resulting from our
baseline model with parameters listed in Table \ref{baselinepars}. }
\label{lcs}
\end{figure}

The snap-shot SEDs in Fig. \ref{SEDs} illustrate that the high-energy end
of the synchrotron peak, emitted directly behind the forward and reverse
shock fronts, remains essentially unaffected as long as the observer receives
synchrotron emission from the shocks still being located within the shells.
As the shocks propagate, an increasingly larger region of the shells is 
energized with particles having longer time to cool. Therefore, the synchrotron
spectrum extends to progressively lower frequencies. As the observer sees the
shock regions leaving the shells, the highest-energy electrons rapidly cool 
off, leading to a loss of the high-frequency synchrotron emission. In the SSC 
component, the light-travel time delay leads to a delayed response of the 
SSC emission with respect to the synchrotron emission, with slightly
cooled electrons still being able to efficiently Thomson scatter synchrotron
photons up to $\gamma$-ray energies.

Fig. \ref{lcs} shows the light curves in the optical (R-band), X-rays
(1 keV), 
and high-energy (HE) $\gamma$-rays in the
Fermi range (100~MeV
and 1~GeV) resulting from our baseline model simulation. The
synchrotron-dominated optical and the EC-dominated HE $\gamma$-ray
light curves exhibit similar shapes, dominated by a rapid onset of
the emission, a continued gradual build-up as long as the shocks are
located within the shells, and a rapid decay dominated by radiative
cooling. The time scale of this decay is inversely proportional to
the characteristic electron energy responsible for the respective
emission. This explains the more rapid decay of the 1~GeV light
curve compared to the 100~MeV one. The X-ray light curve, dominated
by SSC emission, exhibits a much more gradual decay due to the
broad-band convolution of electron and synchrotron photon energies
involved to produce emission at any given frequency. At $\sim 1.6
\times 10^4$~s, the optical emission begins to be dominated by
low-frequency SSC emission characterized by a slow, gradual decay. 

We point out that generally, blazars exhibit a low intensity of 
quiescent emission throughout the electromagnetic spectrum, which
will prevent the observed light curves to reach the very low emission
levels predicted for the beginning and very late times of our simulation.
Assuming that this quiescent level of emission exhibits only moderate
variability, it will lead to more moderate variability amplitudes, but
will not affect the conclusions about cross correlations and time lags
presented in the following.

\begin{figure}[ht]
\plotone{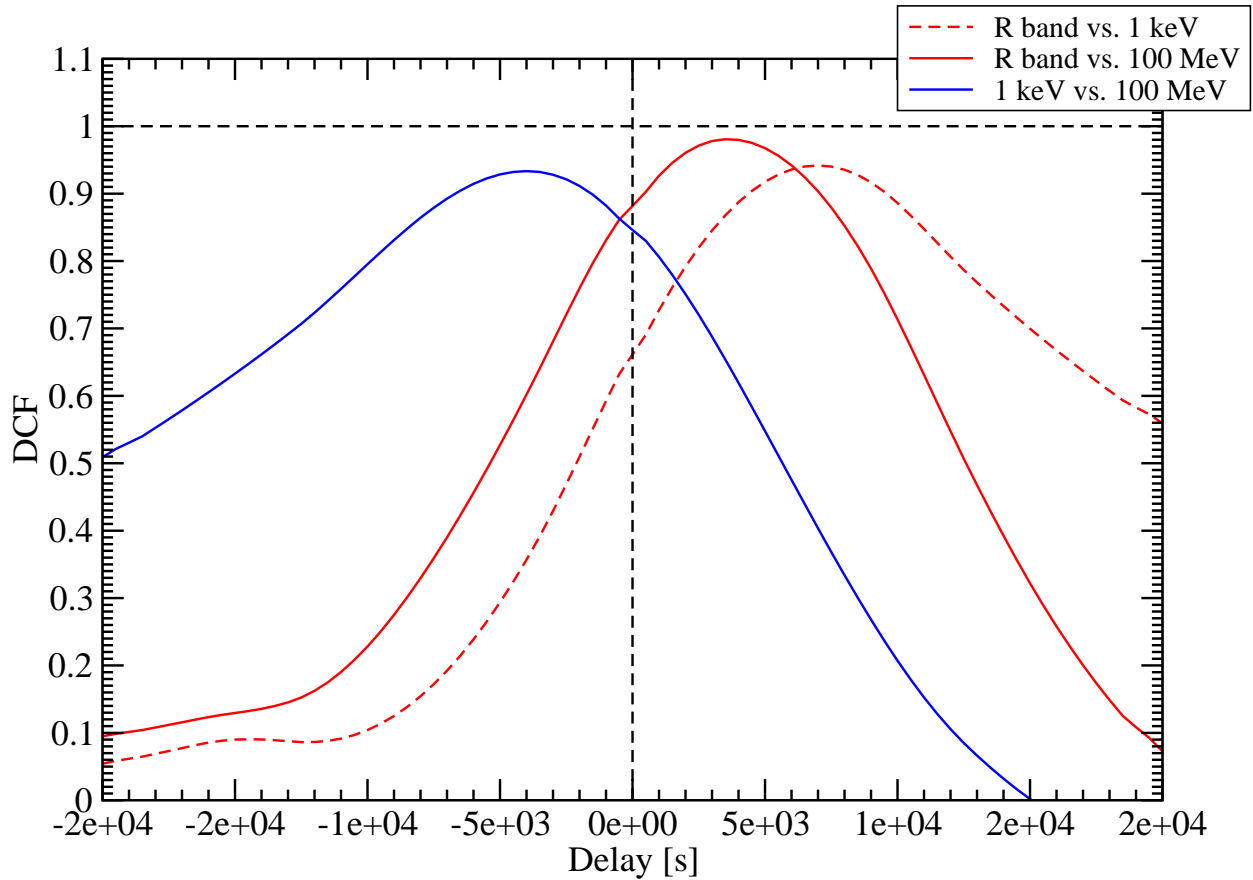}
\caption{Discrete correlation functions between all energy bands for
which light curves are plotted in Figure \ref{lcs}. A positive delay
of "band 1 vs. band 2" indicates a lead of band 1 before band 2.}
\label{DCFs}
\end{figure}

Even though some features of the SEDs and light curves in Figures 
\ref{SEDs} 
and \ref{lcs} are artifacts of our simplified 
analytical treatment of the time- and space-dependent particle distributions 
and the radiation processes, our results are expected to capture the salient 
spectral and light curve features as they would result from a more detailed 
numerical treatment. In particular, comparison is to be made with cross-correlation 
and time lag features that can realistically be extracted from unevenly sampled 
observational data. This is routinely done using the DCF of \cite{ek88}.
Thefore, for direct comparison with observational results, we apply the
same formalism to our simulated light curves. 
The DCFs between the optical,
X-ray and 100~MeV light curves shown in Figure \ref{lcs}, are plotted in
Figure \ref{DCFs}. 

Figure \ref{DCFs} illustrates that the more gradual decay of the
(SSC-dominated) X-ray light curve with respect to the optical one 
translates into an optical lead before the X-rays by about 2~hr. 
The 100~MeV (EC) emission is produced by electrons of lower energy
than the R-band optical synchrotron emission. This leads to a more
gradual light curve decay at 100~MeV, translating into a $\gamma$-ray
lag of $\sim$~1 hr behind the optical, and a reduced lag of X-rays
behind the 100~MeV $\gamma$-rays (compared to the X-ray -- optical
lag) by only about $\sim 1$~hr. All bands show a strong correlation
with a DCF peak $> 0.9$. 

The characteristics of the injected electron distribution are evaluated
according to Eqs. (\ref{gamma1}) and (\ref{gamma2}). In our baseline
model, these are $\gamma_{1,f} = 130$, $\gamma_{1,r} = 350$, and 
$\gamma_{2; f,r} = 4.6 \times 10^4$. The magnetic field, according 
to Eq. \ref{Bfield} is 1.0~G in both shocked regions. We note that
the inferred values of $\gamma_1$ are substantially higher than 
the characteristic values of $\gamma_1 \sim $~a~few inferred by 
\cite{ghisellini09} and \cite{sikora09}. In principle, similar values
could also be achieved with different parameter choices in our
simulations. However, in order to still produce a synchrotron peak 
in the same range as observed in FSRQs, a substantially higher
magnetic field ($B \gtrsim 10$~G) would be required. Therefore, 
if the model explored here is applicable to FSRQs and LBLs with
parameters close to the ones chosen in our parameter study, one
would infer a substantially smaller total number of nonthermal
electrons and hence a smaller total jet power than found in
\cite{ghisellini09} and \cite{sikora09}.

Starting from our baseline model discussed above, we are now investigating
the influence of the various parameters listed in Table \ref{baselinepars}
on the SED and light curve features. For each set of parameters, we evaluate
the time-averaged SED over an integration time of 30~ksec to find the location
of the synchrotron and Compton peaks as well as the Compton dominance,
defined as $CD = \nu F_{\nu}^{\rm IC} / \nu F_{\nu}^{\rm sy}$. For each 
simulation, we calculate the DCFs between the optical (R-band), X-ray
(1~keV), and HE $\gamma$-ray (100~MeV) light curves, and find the predicted 
time delay and peak value of the DCF. 

Specifically, we explore variations of the following parameters and their
influence on the resulting time-averaged SED, inter-band time lags and
DCF peaks: 

\begin{itemize}

\item{The external radiation energy density, $u_{\rm ext}$}

\item{The electron equipartition fraction, $\epsilon_e$}

\item{The magnetic-field equipartition fraction, $\epsilon_B$}

\item{The fraction of electrons accelerated, $\zeta_e$}

\item{The electron acceleration time scale parameter, $a_{\rm acc}$}

\item{The electron injection spectral index, $q$}

\item{The time scales of relativistic shell ejection, $\Delta t_{\rm a,b}$}

\item{The relative Lorentz factors of the colliding shells, $\Gamma_{\rm rel}
\approx (1/2) (\Gamma_a/\Gamma_b + \Gamma_b/\Gamma_a)$}

\item{The kinetic power of the faster shell, $L_b$}

\item{The radius of the shells, $R$}

\end{itemize}

Each parameter is varied individually, leaving all other parameters constant
at the value of our baseline model (Table \ref{baselinepars}). In the following, 
we discuss the influence of each individual parameter mentioned above, on the 
SED and DCF characteristics.

\begin{figure}[ht]
\plotone{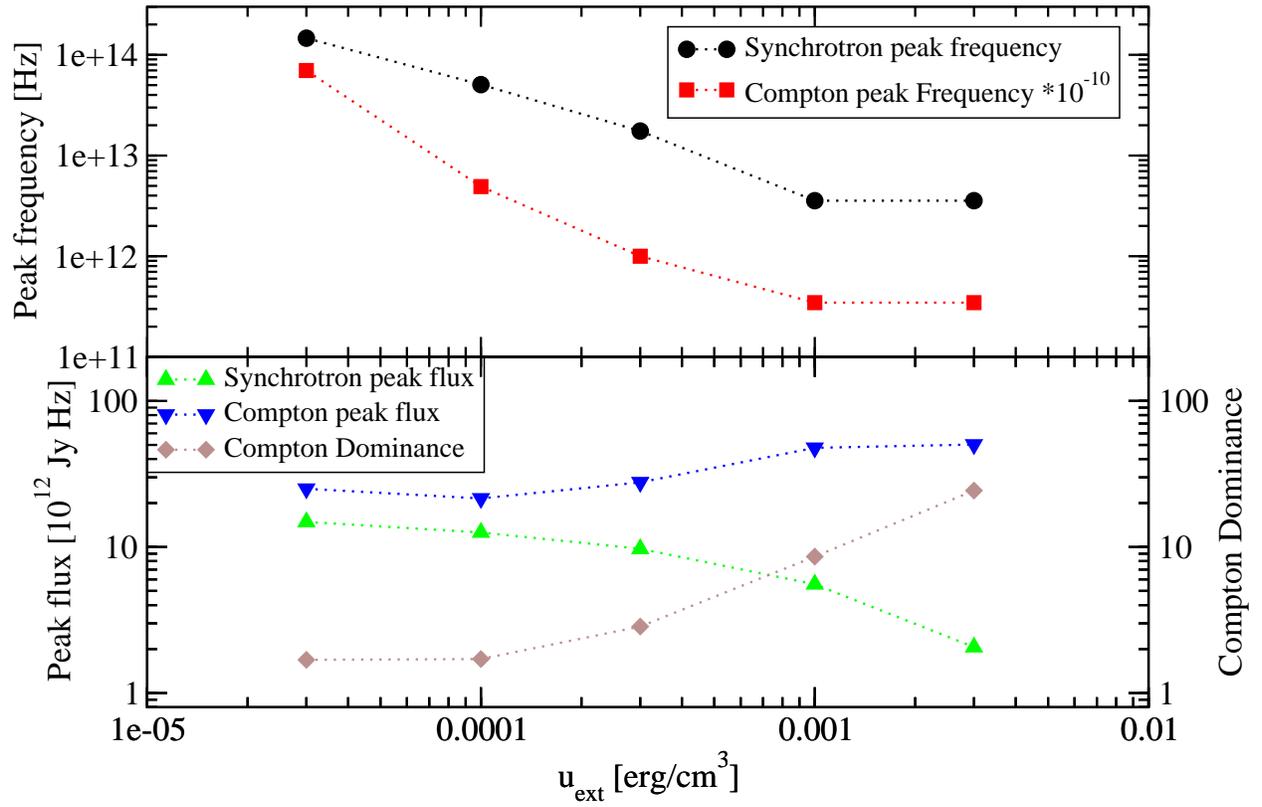}
\caption{Characteristics of the time-averaged SED as a function of the external 
radiation energy density $u_{\rm ext}$. }
\label{uSED}
\end{figure}

\subsection{\label{uext}Variations of $u_{\rm ext}$}

An increasing external radiation energy density leads to a transition from SSC to
EC dominated $\gamma$-ray production and overall more rapid electron cooling. The
consequences in the SED are decreasing synchrotron and EC peak frequencies, along
with a decreasing synchrotron peak flux and an increasing Compton peak flux (see 
Fig. \ref{uSED}). This shift in SED characteristics \citep{fos98} reflects the 
gradual transition from BL-Lacs to FSRQs along the blazar sequence as suggested 
by \cite{ghisellini98}.
The dependence of the external field energy density shown 
in Fig. \ref{uSED} would only apply 
to the FSRQs and LBLs, which have peak synchrotron frequencies between $10^{13}$ 
and $10^{14}$ Hz. In order to extend this to XBLs and the blazar sequence,
it may 
also be necessary to posit a correlation between jet power and the external 
broad-line region density \citep{bd02,gt08}.

\begin{figure}[ht]
\plotone{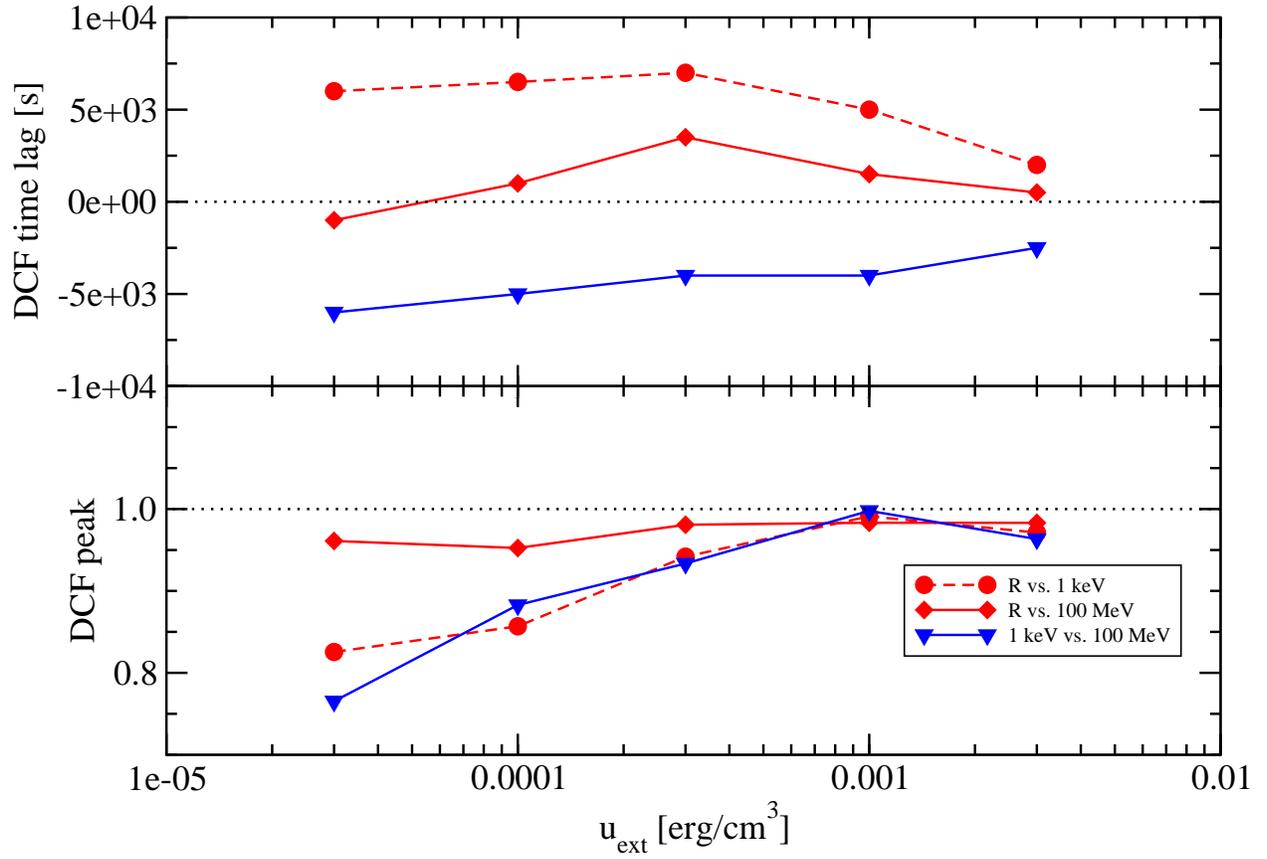}
\caption{Characteristics of the DCFs as a function of the external radiation energy
density $u_{\rm ext}$. }
\label{uDCF}
\end{figure}

Fig. \ref{uDCF} illustrates the 
dependence of time lags and DCF peak values on the external radiation energy 
density. For high values of $u_{\rm ext}$ ($\gtrsim 3 \times 10^{-4}$~erg~cm$^{-3}$),
the more rapid cooling with incresing $u_{\rm ext}$ leads to generally decreasing
time lags. For smaller values of $u_{\rm ext}$, we begin to see the effects of
SSC emission becoming significant in the $\gamma$-ray regime. Most notably, this
ultimately leads to an inversion of the optical -- HE $\gamma$-ray lead into a
lag of optical behind HE $\gamma$-rays for very low $u_{\rm ext}$. The reason
for this is that the SSC $\gamma$-rays at 100~MeV are produced by electrons
of higher energy than those emitting synchrotron radiation in the R band.

All three bands remain generally well correlated (DCF peak $\gtrsim 0.75$),
with only a weak trend towards lower correlation quality for SSC dominated
high-energy emission.

\begin{figure}[ht]
\plotone{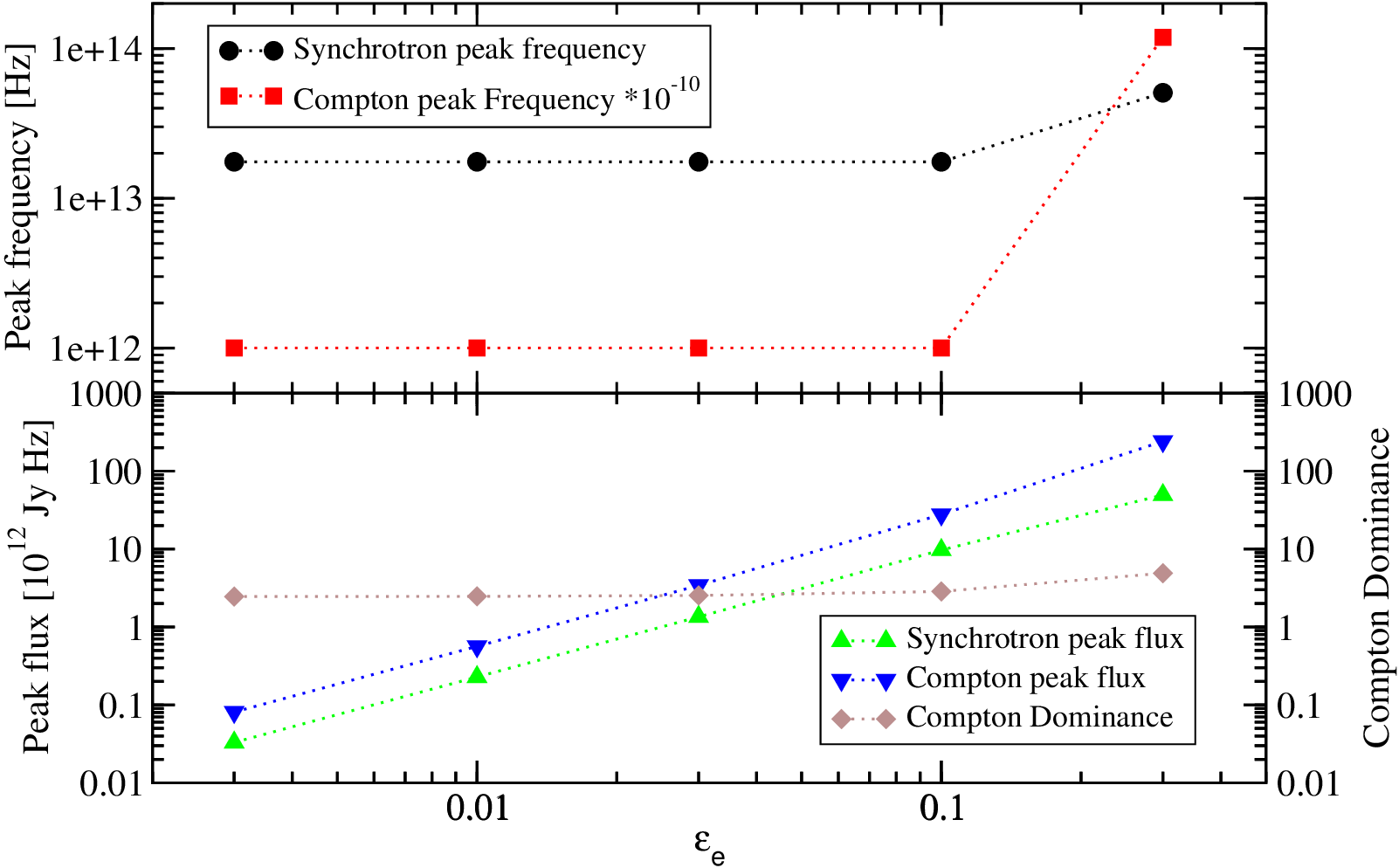}
\caption{Characteristics of the SEDs as a function of the electron equipartition
fraction $\epsilon_e$. }
\label{eeSED}
\end{figure}

\subsection{\label{epse}Variations of $\epsilon_e$}

With increasing electron equipartition fraction $\epsilon_e$, the total energy input
into relativistic electrons increases, along with an increase in the low-energy cutoff
of the injected electron distribution. The increased electron energy density leads to
an increasing radiative output in both spectral components (synchrotron and Compton)
as well as an increasing fraction of SSC to synchrotron emission. 
Since the HE $\gamma$-ray emission remains EC dominated for our
parameters, the Compton dominance remains essentially unchanged.
The peak frequencies of the two spectral components
remain essentially unchanged for realistic values of $\epsilon_e \lesssim 0.1$ 
(see Figure \ref{eeSED}). 

\begin{figure}[ht]
\plotone{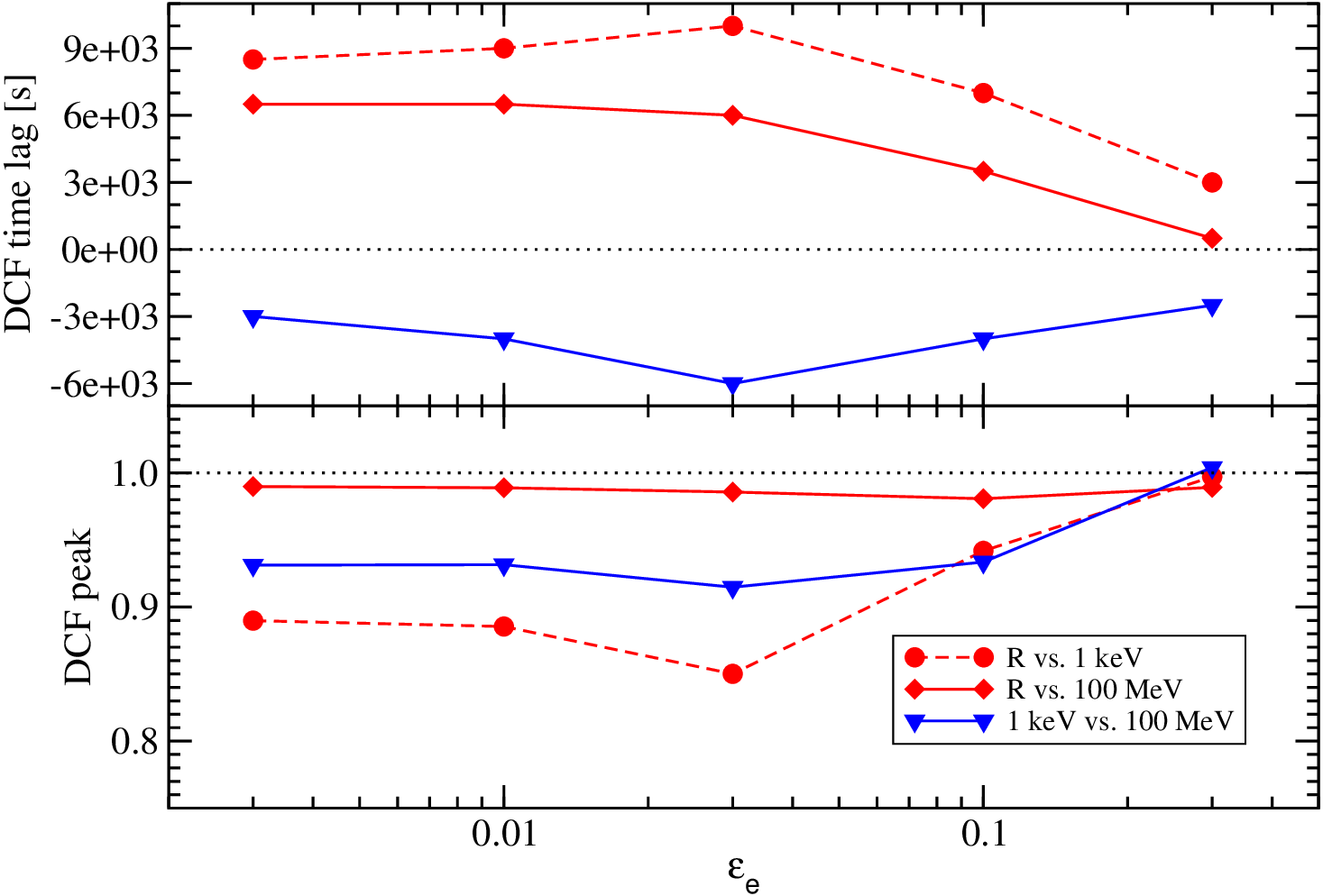}
\caption{Characteristics of the DCFs as a function of the electron equipartition fraction
$\epsilon_e$. }
\label{eeDCF}
\end{figure}

There is an overall weak trend of decreasing inter-band time lags with
increasing $\epsilon_e$, which is related to decreasing electron cooling time scales.
All three bands remain well correlated (DCF peak $\gtrsim 0.85$) irrespective of 
$\epsilon_e$.

\begin{figure}[ht]
\plotone{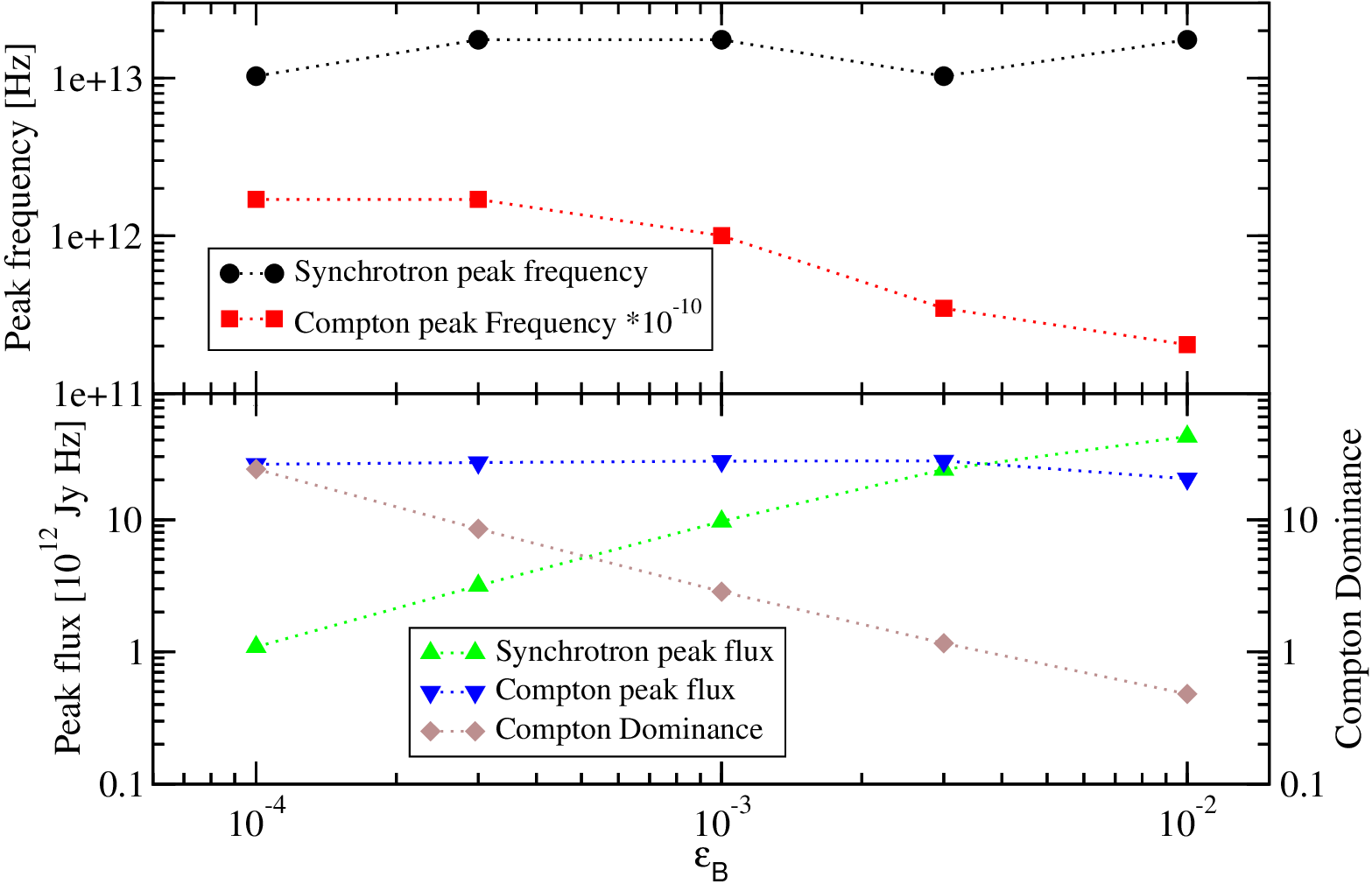}
\caption{Characteristics of the SEDs as a function of the magnetic-field equipartition
fraction $\epsilon_B$. }
\label{eBSED}
\end{figure}

\subsection{\label{epsb}Variations of $\epsilon_B$}

The parameter $\epsilon_B$ regulates the magnetic field. An increasing value of $\epsilon_B$
implies stronger synchrotron cooling.
In the synchrotron frequency range, this effect is largely cancelled out by the increase 
in the characteristic synchrotron frequency ($\nu_{\rm sy} \propto \gamma^2 \, B$). Therefore, 
the synchrotron peak frequency remains essentially unchanged as $\epsilon_B$ changes, while 
the Compton peak frequency decreases (see Fig. \ref{eBSED}).

As a consequence of the limited amount of power available to channel into synchrotron and
Compton emission, the Compton peak flux remains almost constant with varying $\epsilon_B$,
while the synchrotron peak flux continuously increases, consequently leading to a decreasing
Compton dominance.

\begin{figure}[ht]
\plotone{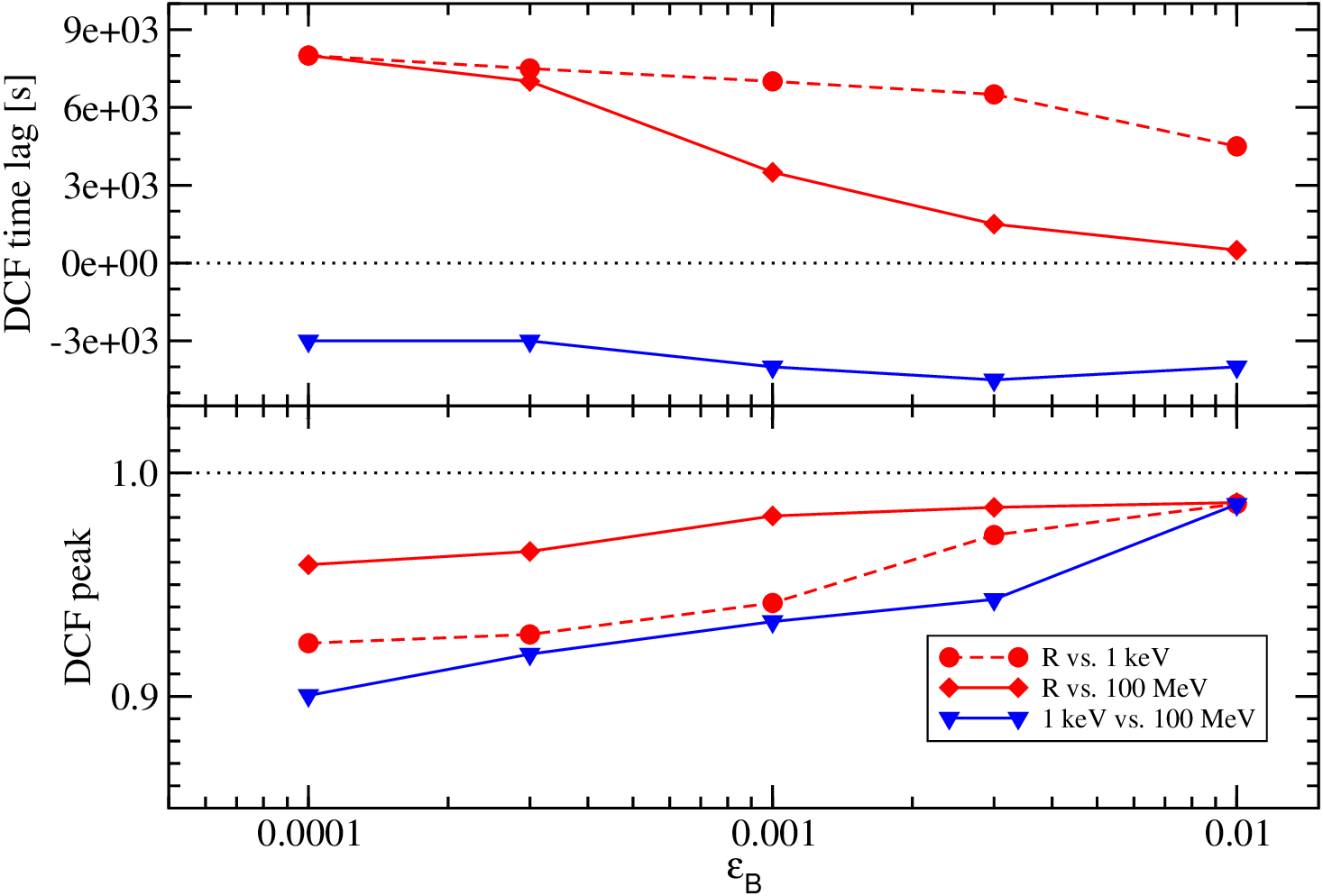}
\caption{Characteristics of the DCFs as a function of the magnetic-field equipartition 
fraction $\epsilon_B$. }
\label{eBDCF}
\end{figure}

In the DCFs (see Figure \ref{eBDCF}), there is a general trend of decreasing time lags and 
increasing quality of correlations with increasing $\epsilon_B$. This is a consequence of
the decreasing radiative cooling time scales 
(with progressively larger contributions from synchrotron and
SSC cooling).

\subsection{\label{zetae}Variations of $\zeta_e$}

Increasing the fraction $\zeta_e$ of electrons accelerated, while keeping the power injected
into relativistic electron (parametrized by $\epsilon_e$) constant, leads to a decreasing
low-energy cutoff of the electron distribution. As $\zeta_e$ increases, an increasing
fraction of the available power in electrons will be distributed at low electron energies,
leading to a decreasing radiative output. Consequently, 
both the synchrotron and Compton
peak frequencies as well as the synchrotron and Compton
peak fluxes decrease with increasing $\zeta_e$. 

Less efficient radiative cooling because of lower average particle energies leads to a
gradual trend of increasing time lags with increasing $\zeta_e$.

\begin{figure}[ht]
\plotone{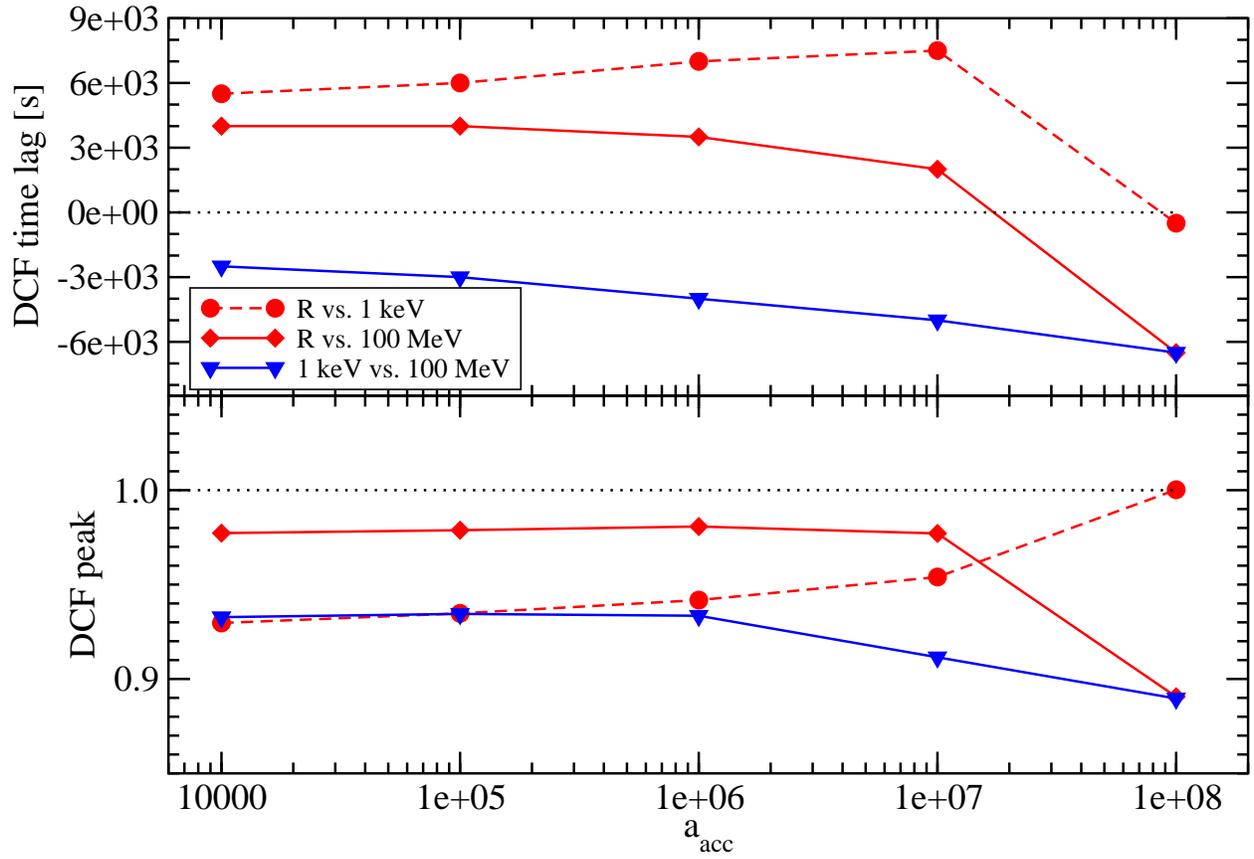}
\caption{Characteristics of the DCFs as a function of the electron acceleration efficiency
parameter $a_{\rm acc}$. }
\label{aaccDCF}
\end{figure}

\subsection{\label{aacc}Variations of $a_{\rm acc}$}

The electron acceleration efficiency parameter $a_{\rm acc}$ regulates the high-energy
cutoff of the injected electron population via Eq. \ref{gamma2}. A lower value of 
$a_{\rm acc}$ implies a larger value of $\gamma_2$. Changes in $a_{\rm acc}$ (and 
hence $\gamma_2$) have a negligible influence on the location of the sychrotron and 
Compton peaks, both in frequency and flux (for electron spectral indices $q > 2$). 
However, the value of $a_{\rm acc}$ (and hence, $\gamma_2$) determines whether the
synchrotron emission of freshly injected electrons extends into the X-ray regime 
or cuts off at optical-UV frequencies. 
As $a_{\rm acc}$ increases, the synchrotron emission cuts off at progressively
lower frequencies. Specifically, this leads to an increasingly rapid decline of the
R-band emission after the shocks have broken out of the shells. As long as the 
synchrotron emission of freshly emitted electrons does extend beyond the R-band,
this leads to an increasing R-band lead before the X-ray emission. For a very high
value of $a_{\rm acc}$ ($10^8$ in our simulations), even the synchrotron emission 
of freshly injected electrons cuts off at frequencies close the R-band, and the R-band 
flux is dominated by low-frequency SSC emission very early on. This reverses the R-band 
vs. X-ray lead into a lag. A similar argument applies to the dependence of the 100~MeV 
$\gamma$-ray emission with respect to X-rays and optical.

All light curves correlate well with DCF peakds $\gtrsim 0.9$, with no substantial
dependence on $a_{\rm acc}$.

\begin{figure}[ht]
\plotone{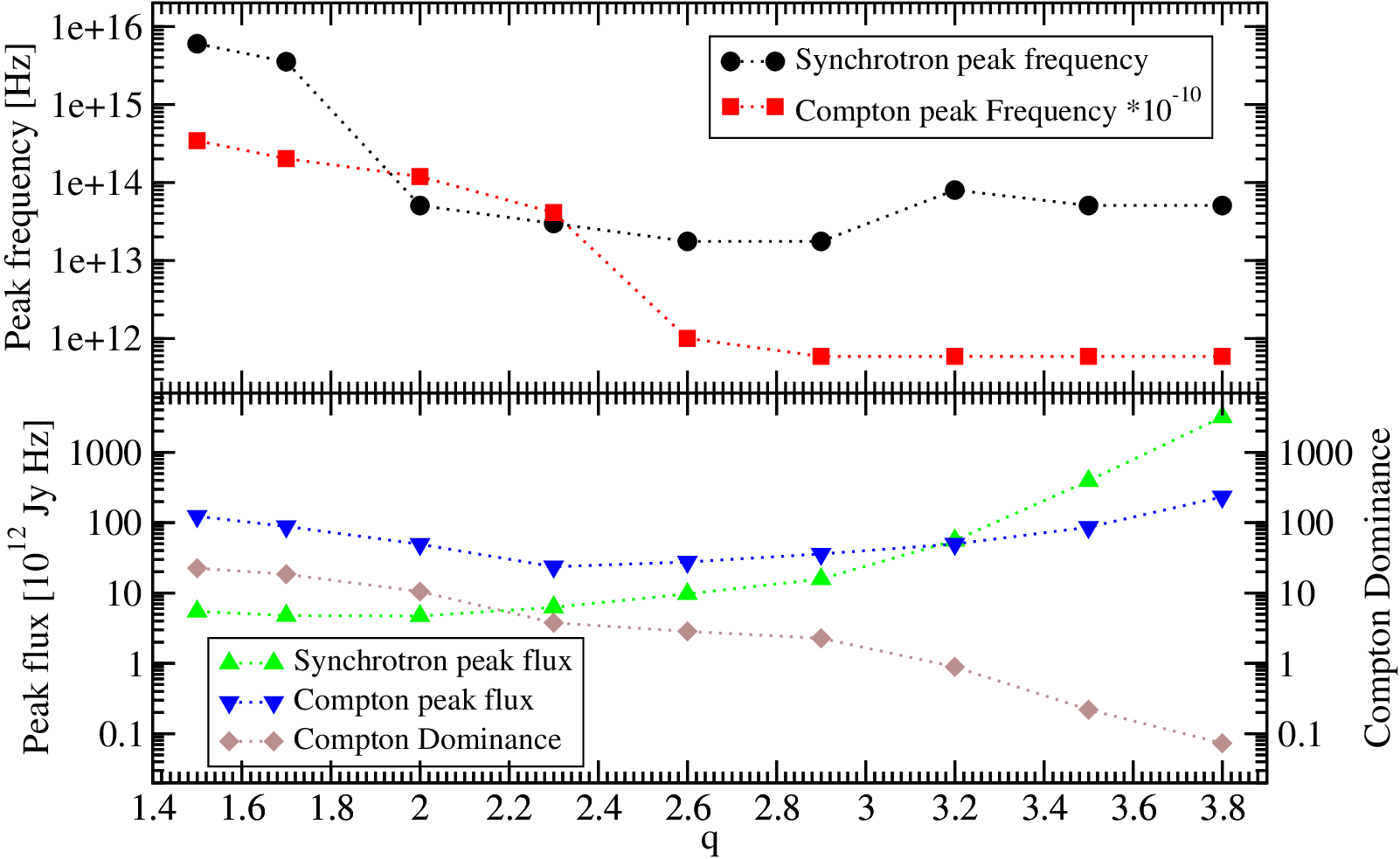}
\caption{Characteristics of the SEDs as a function of the electron injection spectral
index $q$. }
\label{qSED}
\end{figure}

\subsection{\label{qvar}Variations of $q$}

The electron spectral index at the time of injection, $q$, obviously directly determines
the spectral shape of the synchrotron and Compton emissions. 
In particular, for hard injection
spectra ($q \lesssim 2$),
the peak frequencies are determined primarily by the energy of 
the highest-energy electrons.
As we choose a steeper injection electron spectrum, the peak 
shifts to a position
where it is predominantly determined by the low-energy cut-off 
$\gamma_1$ and thus only weakly dependent on the electron injection
index (see Figure 
\ref{qSED}).

Generally, a harder electron injection spectrum (lower $q$) implies that a larger fraction 
of the energy transferred to electrons is stored in high-energy electrons. This leads to a 
monotonic increase of the Compton dominance with decreasing $q$ (harder
electron spectrum).

\begin{figure}[ht]
\plotone{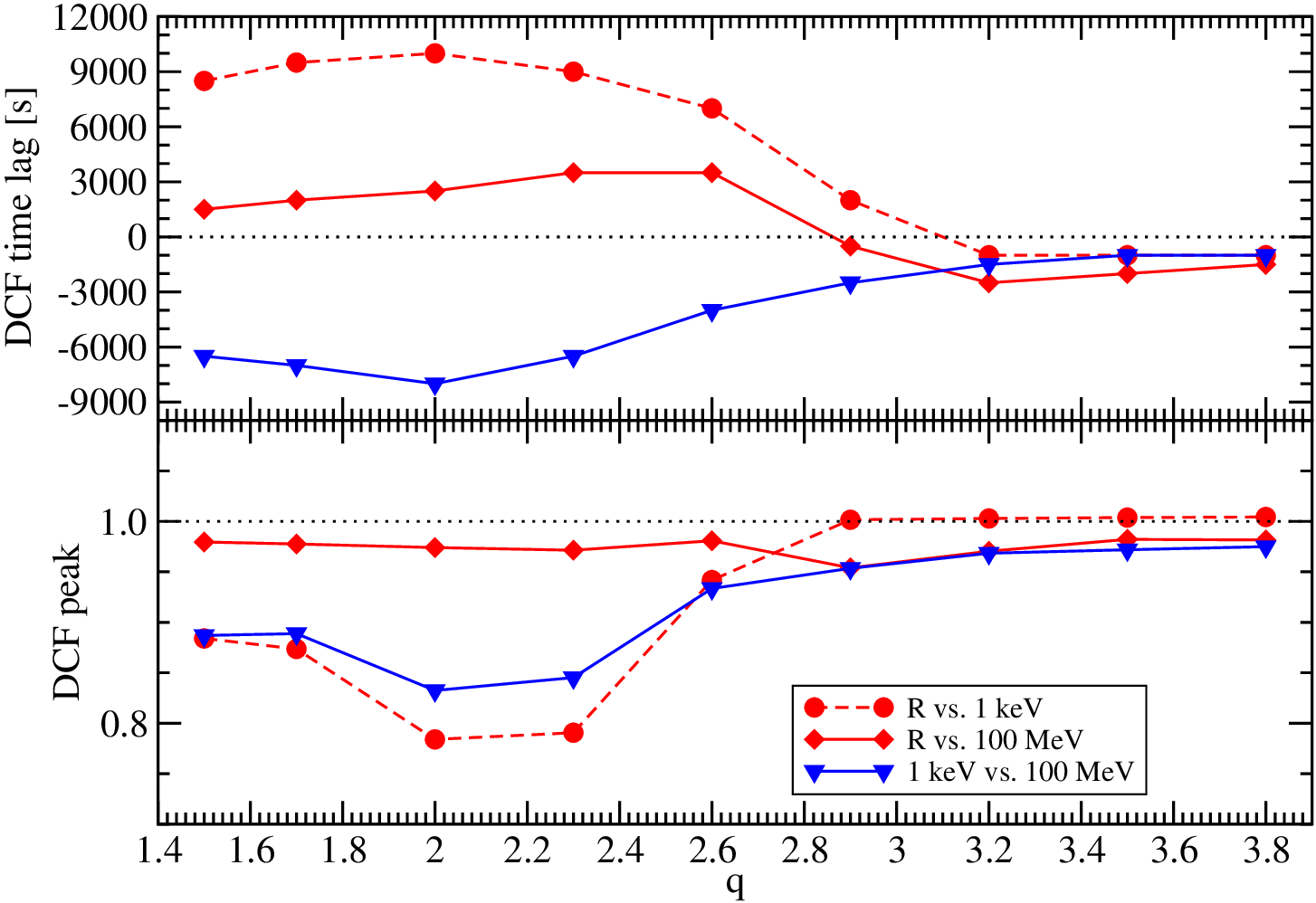}
\caption{Characteristics of the DCFs as a function of the electron injection spectral
index $q$. }
\label{qDCF}
\end{figure}

Generally, all frequency bands correlate well with DCF peak values $\gtrsim 0.8$.
There is a slight trend of de-correlation of the HE $\gamma$-rays with optical and
X-rays for hard injection spectra ($q \lesssim 2.3$). The increasing quality of
correlation with softening injection spectra goes in tandem with decreasing absolute
values of the time lags. Notably, for very steep injection spectra ($q \ge 3$), the
instantaneous synchrotron emission in the R band has a very steep spectral index so
that a substantial fraction of the R-band flux is contributed by very-low-frequency 
SSC emission. This reverses the R-band lead observed for harder injection spectra
into an R-band lag behind X-rays and HE $\gamma$-rays.

\subsection{\label{dtvar}Variations of $\Delta t_{\rm a,b}$}

The shell ejection time scale determines the width of the shells. This in turn, regulates
the shock propagation time and thereby the dynamical time scale of the shocked emission
region. Furthermore, since we left the ejection power constant, a longer ejection time
corresponds to a larger total kinetic energy deposited into the shells. This latter effect
causes the overall radiative power output to increase with $\Delta t_{\rm a,b}$. The Compton 
peak is dominated by EC, so that the Compton dominance remains essentially unchanged. 
The values of the peak frequencies show only a weak increasing trend with increasing
$\Delta t_{\rm a,b}$.

\begin{figure}[ht]
\plotone{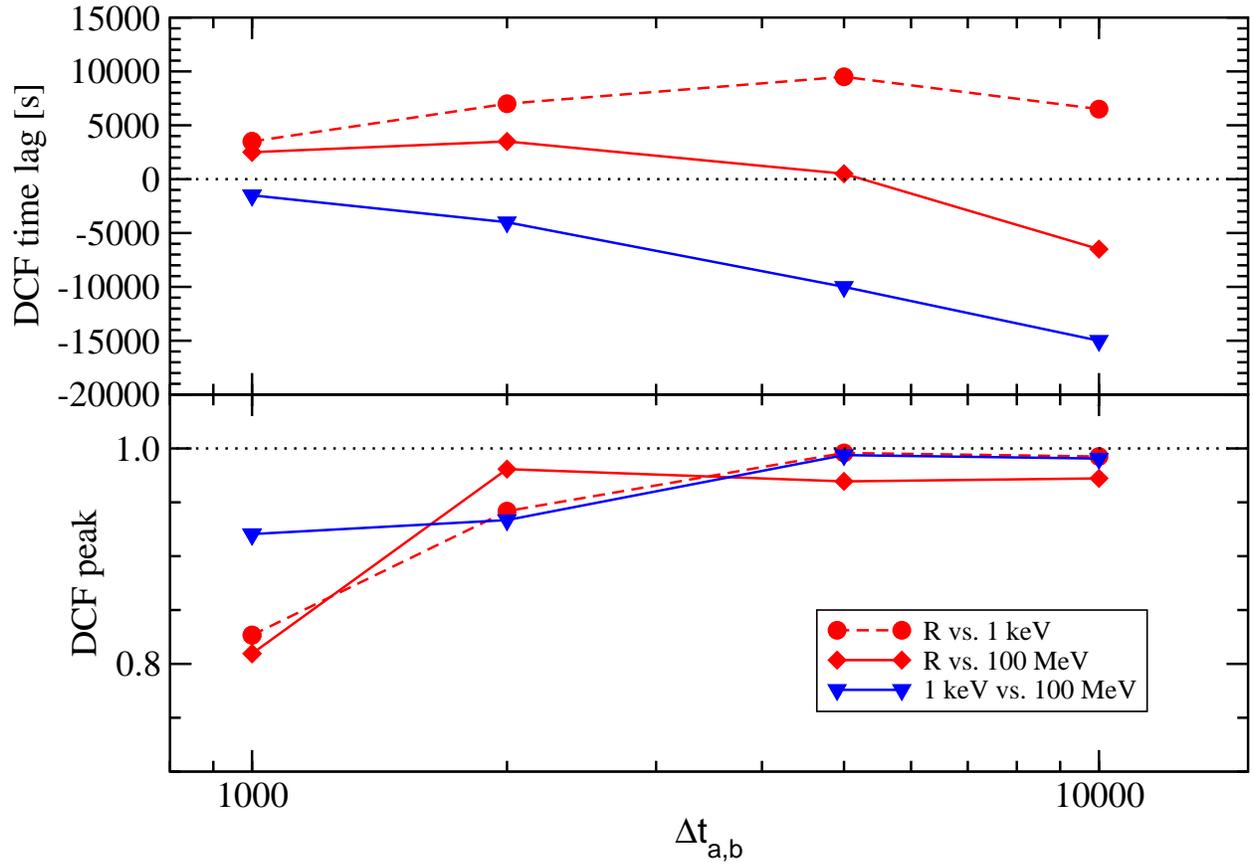}
\caption{Characteristics of the DCFs as a function of the shell ejection time scales,
$\Delta t_{\rm a,b}$ }
\label{dtDCF}
\end{figure}

The most notable trend in the DCF features is a 
reversal of the lag of R behind HE $\gamma$-rays
(100~MeV) into a lead for 
long shock propagation time scales, in tandem with an increasing X-ray lag behind
HE $\gamma$-rays. This is due to the fact that both X-rays and R-band are dominated,
at least at late times, by slowly decaying SSC emission. As the shocked regions are
more extended, SSC emission persists for a longer time due to light-travel time
delays pertinent to the SSC emission. In contrast, the decay of the EC-dominated
HE $\gamma$-ray emission is determined only by the radiative cooling time scale,
which is independent of $\Delta t_{\rm a,b}$ with the assumptions made here. 
It is notable that the time lags are generally not proportional to the dynamical
time of the shock propagation, as one could naively expect.

\begin{figure}[ht]
\plotone{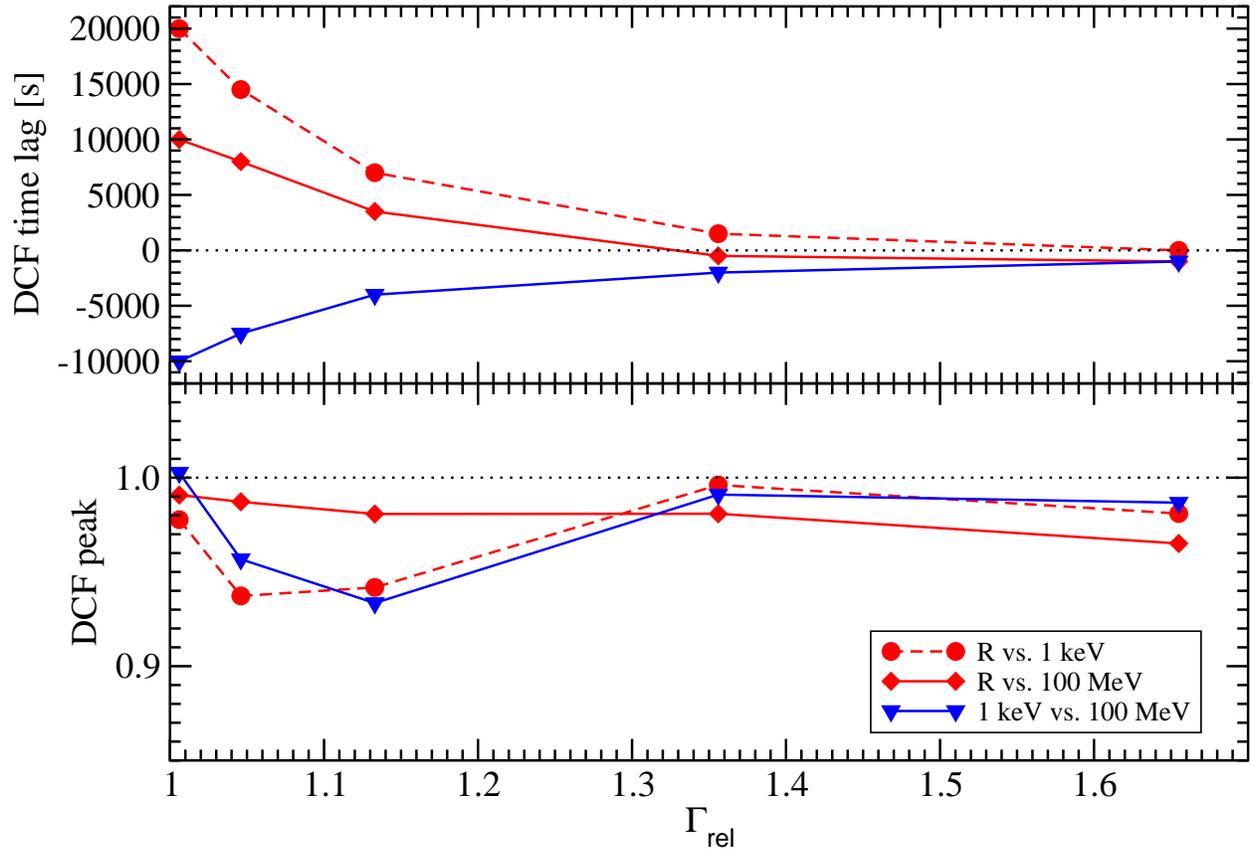}
\caption{Characteristics of the DCFs as a function of the relative Lorentz factor
between the shells, $\Gamma_{\rm rel}$. }
\label{GammarelDCF}
\end{figure}

\subsection{\label{Gammarelvar}Variations of $\Gamma_{\rm rel}$} 

We have explored the effect of an increasing difference in the Lorentz factors of 
the colliding shells, leading to an increasing relative Lorentz factor, $\Gamma_{\rm rel} 
\approx (1/2) \, (\Gamma_a / \Gamma_b + \Gamma_b / \Gamma_a)$, by varying $\Gamma_a$
and $\Gamma_b$ in a way that the resulting shocked-fluid Lorentz factor $\Gamma$
(and hence the Doppler boosting factor $D$) remains unchanged. An increase in 
$\Gamma_{\rm rel}$ drastically increases the radiative efficiency of the shells. 
The increased energy density in the shocked shells will, at the same time, 
increase the magnetic fields and decrease the radiative cooling time scales. 
It will also lead to larger shock propagation Lorentz factors 
$\overline\Gamma_{f,r}$ and hence larger low-energy cut-offs in the electron 
distributions. This leads to a net increase of the peak frequencies of both
radiation components with increasing $\Gamma_{\rm rel}$. The peak fluxes in
both components increase drastically with increasing $\Gamma_{\rm rel}$ with
only a moderate effect on the Compton dominance. 
The decreasing radiative cooling time scales with increasing $\Gamma_{\rm rel}$ 
imposes
a general trend of shortening inter-band time lags with only minor effect 
on the
DCF peak values (see Fig. \ref{GammarelDCF}).

\subsection{\label{Lbvar}Variations of $L_b$}

Varying the kinetic luminosity of one of the shells, keeping its Lorentz factor
unchanged, directly amounts to increasing its particle density. For unchanged
$\epsilon_B$, this will also increase the magnetic field. Both effects lead to
an increasing synchrotron peak flux and a decreasing Compton dominance. Due to
more rapid radiative (synchrotron) cooling, both peak peak frequencies decrease
with increasing $L_b$. 
As the system becomes increasingly synchrotron + SSC
dominated, the light-travel time delays inherent to SSC emission lead to a
general trend of increasing time delays with increasing $L_b$, accompanied
by very moderate changes in the DCF peak 
values.

\subsection{\label{Rvar}Variations of $R$}

Increasing the shell radius implies a smaller particle density and hence a lower
magnetic field. 
Consequently, the synchrotron peak flux and peak frequency decrease with
increasing shell radius, leading to an increasing Compton dominance.

In the DCFs, an increasing radius $R$ leads to a gradually increasing trend of
time lags between R and higher-frequency bands with slightly decreasing DCF peak
values ($\sim 0.95 \to 0.8$ for $R = 3 \times 10^{15}$~cm $\to \, 3 \times 10^{16}$~cm). 
The increasing time lag trend may be related to the increasing synchrotron cooling
time as $R$ increases.

\section{\label{summary}Summary and Conclusions}

We have developed a semi-analytical internal-shock model for blazars. Our model
includes synchrotron, SSC, and external Compton emission, and properly takes into
account light-travel time and shock-propagation effects as well as the space- and
time-dependent relativistic electron distributions in the shocked shell region to
evaluate the time-dependent radiative output. For direct comparison with 
sparsely and unevenly sampled observational data, we evaluated the Discrete
Correlation Functions between several representative wavelengths, namely between
the optical R band, 1 keV X-rays, and high-energy $\gamma$-rays (100~MeV --- Fermi). 

Our calculations were used to characterize inter-band time lags and the quality of 
the correlation, as
represented by the peak value of the DCF. For this calcuation, 
we considered a specific model blazar where spectral variability could be detected
with optical and pointing
X-ray telescopes at the derived flux level. These calculations 
can be extended to apply to brighter blazars where
variability might be detected with 
the Fermi telescope, or otherwise used to model the time-averaged SED observed with 
Fermi. Keeping these caveats in mind, we have studied the influence of variations
of a number of essential model parameters on the resulting SED and DCF characteristics.

We note that our model results apply to FSRQs and LBLs, maybe IBLs, but not HBLs, 
because of the restriction to the Thomson regime. The parameter choices in our model 
are extensive, including $\epsilon_e$ and $\epsilon_B$ parameters,
shell widths, collision 
radii, and jet powers. We have reduced the allowed parameter space by taking equal shell 
powers, with the forward and reverse shocks having equal $\epsilon_e$ and $\epsilon_B$ 
parameters.
By examining the dependence of peak frequency, peak $\nu F_\nu$ flux and 
Compton dominance in Fig. \ref{uSED}, we find a trend in accord with the behavior 
of the blazar sequence \citep{fos98}.

Our results do not directly explain the break in the $\nu F_\nu$ spectra at a few 
GeV
in blazars like 3C 454.3 and AO~0235+164 and other FSRQs and LBLs 
\citep{abd09a,abd09b}, or the relative constancy
of the GeV spectral index 
with flux state. The GeV breaks in several blazar SEDs can not be explained 
as a cooling break since, whenever photon statistics
allow a reliable determination of the break, it is $\Delta\alpha > 0.5$. However, 
a smooth high-energy cutoff could effectively mimic a spectral break, and such 
a cutoff around a few GeV could be produced in our model by a lower high-energy
cutoff of the electron distribution, which could be achieved with a choice of a
larger value for our paramter $a_{\rm acc}$. 

We have not considered a detailed spectral comparison with the impressive 
multiwavelength SEDs now available due to campaigns involving the Fermi 
Gamma Ray Space Telescope \citep{abd09c}. Rather, here we considered a 
parameter study of short timescale variability between optical, X-ray, 
and gamma-ray eneriges.

In most cases, the optical light curves showed substantial 
leads (by $\sim$ a few hours) before X-ray and HE $\gamma$-ray emission. 
However, variations of several parameters (e.g., the external radiation energy 
density $u_{\rm ext}$, the acceleration time scale parameter $a_{\rm acc}$, 
the electron spectral index $q$, and the shell widths) can drastically change 
the amount and even sign of the time delays.
This may explain the lack of a consistent pattern of inter-band time lags 
when comparing multiple observing epochs of the same object \citep[e.g.,][]{hartman01}.

\acknowledgments

MB acknowledges partial support from NASA through INTEGRAL Guest Investigator grant
NNX09AI71G and Fermi Guest Investigator grant NNX09AT82G. The work of CDD is supported
by the Office of Naval Research.

\appendix

\section{\label{synchrotronapp}Time-dependent Synchrotron Spectra}

Since the particle energy distribution of Eq. \ref{ne} explicitly depends
on $\overline t_{\rm em}$, it is convenient to transform the integration
(\ref{nuFnusy}) into an integration over emission time, using 

\begin{equation}
d\overline t_{\rm em} = \left( {\mu \over \Gamma^2 c} - {{\rm sign}(\overline x)
\over \overline\beta_{f,r} \, c} \right) \, d\overline x.
\label{dtbar}
\end{equation}
As mentioned above, the limits $\overline x_{\rm min,max}$ of Eq. 
\ref{xlimits} will then be augmented by the condition that $n(\gamma_{\rm sy},
\overline t_{\rm em}) > 0$ for a given electron energy $\gamma_{\rm sy}$.
Since $\mu/\Gamma^2 < 1/\overline\beta_{f,r}$ in the parameter range 
relevant to internal shocks in blazars ($\mu \sim 1$; $\Gamma >> 1$, 
$\overline\beta_{f,r} \sim 1$), the emission times will be $\le \overline t
= t_{\rm obs} (1 + z)/D$ throughout the integration region. Thus, defining 
a time limit $\overline t_f$ corresponding to the limit $x_{\rm max}$ in 
the forward shock region and $\overline t_r$ corresponding to $x_{\rm min}$ 
in the reverse shock region, the integral in Eq. \ref{nuFnusy} will consist 
of two branches,

\begin{equation}
\int\limits_{\overline x_{\rm min}}^{\overline 
x_{\rm max}} n_e \left(\sqrt{\epsilon \, (1 + z) \over b \, D} 
\, , \, \overline t_{\rm x, em}\right) \; d \overline x 
= \left\lbrace
\int\limits_{\overline t_f}^{\overline t} {d\overline t_{\rm x, em} \over 
{1 \over \overline\beta_f \, c} - {\mu \over \Gamma^2 c}} +
\int\limits_{\overline t_r}^{\overline t} {d\overline t_{\rm x, em} \over 
{\mu \over \Gamma^2 c} + {1 \over \overline\beta_r \, c}} \right\rbrace
n_e \left(\sqrt{\epsilon \, (1 + z) \over b \, D} \, , \,
\overline t_{\rm x, em}\right)
\label{tintegration}
\end{equation}
The limits on $\overline x$ in Eq. \ref{xlimits} can then be converted
into limits on $\overline t_{\rm x, em}$ as

\begin{eqnarray}
\overline t_r &=& \max\left\lbrace 0 \; , \; \left( \min\left[ {\overline t
\, \overline\beta_r \, c \over 1 + {\mu \, \overline\beta_r \over \Gamma^2}}
\; , \; \overline{\Delta r}_b \right] \left[ {\mu \over \Gamma^2 c}
+ {1 \over \overline\beta_r \, c} \right] \right) \right\rbrace \cr\cr
\overline t_f &=& \max\left\lbrace 0 \; , \; \left( \min\left[ {\overline t
\, \overline\beta_f \, c \over 1 - {\mu \, \overline\beta_f \over \Gamma^2}}
\; , \; \overline{\Delta r}_a \right] \left[ {\mu \over \Gamma^2 c}
- {1 \over \overline\beta_f \, c} \right] \right) \right\rbrace \cr
\label{tlimits}
\end{eqnarray}
We now evaluate each term in the integral (\ref{tintegration}) separately
for each term in the expression for $n$ in Eq. (\ref{ne}):

\begin{equation}
I_{1r} \equiv { Q_0 \, c \, \gamma^{-q} \over \left( {\mu \over \Gamma^2} 
+ {1 \over \overline\beta_r} \right)}
\int\limits_{\overline t_r}^{\overline t} 
H(\gamma_{\rm up} - \gamma) \, H(\gamma; \gamma_{1,r}, \gamma_c) 
\, \min(\overline t_{\rm x, em}, \overline{\Delta t}_{\rm acc, r}) 
\; d\overline t_{\rm x, em}.
\label{I1r}
\end{equation}
This integral contributes only as long as 

\begin{eqnarray}
\overline t_{\rm x, em} \ge& \overline t_r & {\rm (a)} \cr
\overline t_{\rm x, em} \le& \overline t & {\rm (b)} \cr
\gamma \le& \gamma_c = {1 \over \nu_0 \overline t_{\rm x, em}} & {\rm (c)} \cr
\gamma \le& \gamma_{\rm up} = {1 \over \gamma_{2,r}^{-1} + \nu_0 \, \max(0, 
[\overline t_{\rm x, em} - \overline{\Delta t}_{\rm acc, r}])} & {\rm (d)} \cr
\label{t1r}
\end{eqnarray}
These constraints translate into effective limits of the integral
(\ref{I1r}) of

\begin{eqnarray}
\overline t_{\rm 1r, min} =& \overline t_r \cr
\overline t_{\rm 1r, max} =& \min\left\lbrace \overline t \; , \; {1 \over
\nu_0 \gamma} \; , \; \left( {1 \over \nu_0} \, \left[ {1 \over \gamma}
- {1 \over \gamma_2} \right] + \overline{\Delta t}_{\rm acc, r} \right) 
\right\rbrace \cr
\label{I1rlimits}
\end{eqnarray}
and hence,

$$
I_{1r} = {Q_0 \, c \, \gamma^{-q} \over {\mu \over \Gamma^2} +
{1 \over \overline\beta_r}} \, H(\gamma - \gamma_{\rm 1,r})
$$
\begin{equation}
\times 
\cases{(1/2) (\overline t_{\rm 1r, max}^2 - \overline t_{\rm 1r, min}^2)
& if $\overline t_{\rm 1r, min} < \overline t_{\rm 1r, max} < \overline{\Delta
t}_{\rm acc, r}$ \cr
(1/2) (\overline{\Delta t}_{\rm acc, r}^2 - \overline t_{\rm 1r, min}^2)
+ \overline{\Delta t}_{\rm acc, r} \, (\overline t_{\rm 1r, max} -
\overline{\Delta t}_{\rm acc, r}) & if $\overline t_{\rm 1r, min} \le
\overline{\Delta t}_{\rm acc, r} \le \overline t_{\rm 1r, max}$ \cr
\overline{\Delta t}_{\rm acc, r} \, (\overline t_{\rm 1r, max} -
\overline t_{\rm 1r, min}) & if $\overline{\Delta t}_{\rm acc, r} \le
\overline t_{\rm 1r, min} \le \overline t_{\rm 1r, max}$ \cr}
\label{I1rsolution}
\end{equation}

The forward-shock contribution of the first term is 

\begin{equation}
I_{1f} \equiv { Q_0 \, c \, \gamma^{-q} \over \left( {1 \over 
\overline\beta_f} - {\mu \over \Gamma^2} \right)}
\, \int\limits_{\overline t_f}^{\overline t} 
H(\gamma_{\rm up} - \gamma) \, H(\gamma; \gamma_{1,r}, \gamma_c) 
\, \min(\overline t_{\rm x, em}, \overline{\Delta t}_{\rm acc, r}) 
\; d\overline t_{\rm x, em}.
\label{I1f}
\end{equation}
Analogous to the previous integral, the effective limits of this
integration are

\begin{eqnarray}
\overline t_{\rm 1f, min} =& \overline t_f \cr
\overline t_{\rm 1f, max} =& \min\left\lbrace \overline t \; , \; {1 \over
\nu_0 \gamma} \; , \; \left( {1 \over \nu_0} \, \left[ {1 \over \gamma}
- {1 \over \gamma_2} \right] + \overline{\Delta t}_{\rm acc, f} \right) 
\right\rbrace \cr
\label{I1flimits}
\end{eqnarray}
and hence,

$$
I_{1f} = {Q_0 \, c \, \gamma^{-q} \over {1 \over \overline\beta_f} -
{\mu \over \Gamma^2}} \, H(\gamma - \gamma_{\rm 1,f})
$$
\begin{equation}
\times 
\cases{(1/2) (\overline t_{\rm 1f, max}^2 - \overline t_{\rm 1f, min}^2)
& if $\overline t_{\rm 1f, min} < \overline t_{\rm 1f, max} < \overline{\Delta
t}_{\rm acc, f}$ \cr
(1/2) (\overline{\Delta t}_{\rm acc, f}^2 - \overline t_{\rm 1f, min}^2)
+ \overline{\Delta t}_{\rm acc, f} \, (\overline t_{\rm 1f, max} -
\overline{\Delta t}_{\rm acc, f}) & if $\overline t_{\rm 1f, min} \le
\overline{\Delta t}_{\rm acc, f} \le \overline t_{\rm 1f, max}$ \cr
\overline{\Delta t}_{\rm acc, f} \, (\overline t_{\rm 1f, max} -
\overline t_{\rm 1f, min}) & if $\overline{\Delta t}_{\rm acc, f} \le
\overline t_{\rm 1f, min} \le \overline t_{\rm 1f, max}$ \cr}
\label{I1fsolution}
\end{equation}

The reverse-shock contribution from the second term in Eq. \ref{ne}
is

\begin{equation}
I_{\rm 2r} \equiv {Q_0 \, c \, \gamma^{-(q + 1)} \over \left( {\mu \over 
\Gamma^2} + {1 \over \overline\beta_r} \right)} 
\int\limits_{\overline t_r}^{\overline t} 
H(\gamma; \gamma_c, \gamma_{\rm up})  \, H(\gamma - \gamma_{\rm 1, r})
\, {\min(\overline t_{\rm x, em}, \overline{\Delta t}_{\rm acc, r}) 
\over \overline t_{\rm x, em}} \; d\overline t_{\rm x, em}.
\label{I2r}
\end{equation}
with the limits determined by the conditions

\begin{eqnarray}
\overline t_{\rm x, em} \ge& \overline t_r & {\rm (a)} \cr
\overline t_{\rm x, em} \le& \overline t & {\rm (b)} \cr
\gamma \ge& \gamma_c = {1 \over \nu_0 \overline t_{\rm x, em}} & {\rm (c)} \cr
\gamma \le& \gamma_{\rm up} = {1 \over \gamma_{2,r}^{-1} + \nu_0 \, \max(0, 
[\overline t_{\rm x, em} - \overline{\Delta t}_{\rm acc, r}])} & {\rm (d)} \cr
\label{t2r}
\end{eqnarray}
These constraints translate into effective limits of the integral
(\ref{I2r}) of

\begin{eqnarray}
\overline t_{\rm 2r, min} =& \max\left\lbrace \overline t_r \; , \;
{1 \over \nu_0 \, \gamma} \right\rbrace \cr
\overline t_{\rm 2r, max} =& \min\left\lbrace \overline t \; , \; 
\left( {1 \over \nu_0} \, \left[ {1 \over \gamma}
- {1 \over \gamma_{2, r}} \right] + \overline{\Delta t}_{\rm acc, r} \right) 
\right\rbrace \cr
\label{I2rlimits}
\end{eqnarray}
and hence,

$$
I_{\rm 2r} \equiv {Q_0 \, c \, \gamma^{-(q + 1)} \over \left( {\mu \over 
\Gamma^2} + {1 \over \overline\beta_r} \right)} \, H(\gamma - \gamma_{\rm 1, r})
$$
\begin{equation}
\times 
\cases{
\overline t_{\rm 2r, max} - \overline t_{\rm 2r, min}
& if $\overline t_{\rm 2r, min} < \overline t_{\rm 2r, max} < \overline{\Delta
t}_{\rm acc, r}$ \cr
\overline{\Delta t}_{\rm acc, r} - \overline t_{\rm 2r, min}
+ \overline{\Delta t}_{\rm acc, r} \, \ln\left( { \overline t_{\rm 2r, max} 
\over \overline{\Delta t}_{\rm acc, r}} \right) & if $\overline t_{\rm 2r, min} \le
\overline{\Delta t}_{\rm acc, r} \le \overline t_{\rm 2r, max}$ \cr
\overline{\Delta t}_{\rm acc, r} \, \ln\left( {\overline t_{\rm 2r, max} \over
\overline t_{\rm 2r, min}} \right) & if $\overline{\Delta t}_{\rm acc, r} \le
\overline t_{\rm 2r, min} \le \overline t_{\rm 2r, max}$ \cr}
\label{I2rsolution}
\end{equation}

Analogously, for the forward-shock contribution from the second term in
Eq. \ref{ne}, we have the effective limits

\begin{eqnarray}
\overline t_{\rm 2f, min} =& \max\left\lbrace \overline t_f \; , \;
{1 \over \nu_0 \, \gamma} \right\rbrace \cr
\overline t_{\rm 2f, max} =& \min\left\lbrace \overline t \; , \; 
\left( {1 \over \nu_0} \, \left[ {1 \over \gamma}
- {1 \over \gamma_{\rm 2, f}} \right] + \overline{\Delta t}_{\rm acc, f} \right) 
\right\rbrace \cr
\label{I2flimits}
\end{eqnarray}
and hence,

$$
I_{\rm 2f} \equiv {Q_0 \, c \, \gamma^{-(q + 1)} \over \left( {1 \over \overline\beta_f} 
- {\mu \over \Gamma^2} \right) } \, H(\gamma - \gamma_{\rm 1, f})
$$
\begin{equation}
\times 
\cases{
\overline t_{\rm 2f, max} - \overline t_{\rm 2f, min}
& if $\overline t_{\rm 2f, min} < \overline t_{\rm 2f, max} < \overline{\Delta
t}_{\rm acc, f}$ \cr
\overline{\Delta t}_{\rm acc, f} - \overline t_{\rm 2f, min}
+ \overline{\Delta t}_{\rm acc, f} \, \ln\left( { \overline t_{\rm 2f, max} 
\over \overline{\Delta t}_{\rm acc, f}} \right) & if $\overline t_{\rm 2f, min} \le
\overline{\Delta t}_{\rm acc, f} \le \overline t_{\rm 2f, max}$ \cr
\overline{\Delta t}_{\rm acc, f} \, \ln\left( {\overline t_{\rm 2f, max} \over
\overline t_{\rm 2f, min}} \right) & if $\overline{\Delta t}_{\rm acc, f} \le
\overline t_{\rm 2f, min} \le \overline t_{\rm 2f, max}$ \cr}
\label{I2fsolution}
\end{equation}

The third term in Eq. \ref{ne} yields a reverse-shock contribution of
\begin{equation}
I_{\rm 3r} \equiv {Q_0 \, c \, \gamma^{-2} \, \gamma_{\rm 1,r}^{(2 - q)} \over 
\left( {\mu \over \Gamma^2} + {1 \over \overline\beta_r} \right)} 
\int\limits_{\overline t_r}^{\overline t} 
H(\gamma; \gamma_{\rm min}, \gamma_{\rm 1, r}) \; d\overline t_{\rm x, em}.
\label{I3r}
\end{equation}
with the limits determined by the conditions

\begin{eqnarray}
\overline t_{\rm x, em} \ge& \overline t_r & {\rm (a)} \cr
\overline t_{\rm x, em} \le& \overline t & {\rm (b)} \cr
\gamma \ge& \gamma_{\rm min} = {1 \over \gamma_{\rm 1,r}^{-1} + \nu_0 \overline 
t_{\rm x, em}} & {\rm (c)} \cr
\gamma \le& \gamma_{\rm max} & {\rm (d)} \cr
\gamma \le& \gamma_{\rm 1,f} & {\rm (e)} \cr
\label{t3r}
\end{eqnarray}
These constraints translate into effective limits of the integral
(\ref{I3r}) of

\begin{eqnarray}
\overline t_{\rm 3r, min} =& \max\left\lbrace \overline t_r \; , \;
{1 \over \nu_0} \left( {1 \over \gamma} - {1 \over \gamma_{\rm 1,r}} \right) 
\right\rbrace \cr
\overline t_{\rm 3r, max} =& \min\left\lbrace \overline t \; , \; 
\left( {1 \over \nu_0} \, \left[ {1 \over \gamma}
- {1 \over \gamma_{2, r}} \right] + \overline{\Delta t}_{\rm acc, r} \right) 
\right\rbrace \cr
\label{I3rlimits}
\end{eqnarray}
and hence,

\begin{equation}
I_{\rm 3r} \equiv {Q_0 \, c \, \gamma^{-2} \, \gamma_{\rm 1,r}^{(2 - q)} \over 
\left( {\mu \over \Gamma^2} + {1 \over \overline\beta_r} \right)} \times 
(\overline t_{\rm 3r, max} - \overline t_{\rm 3r, min})
\label{I3rsolution}
\end{equation}

Finally, the contribution of the third term in Eq. \ref{ne} from the forward
shock is

\begin{equation}
I_{\rm 3f} \equiv {Q_0 \, c \, \gamma^{-2} \, \gamma_{\rm 1,r}^{(2 - q)} \over 
\left( {1 \over \overline\beta_r} - {\mu \over \Gamma^2} \right)} \times 
(\overline t_{\rm 3f, max} - \overline t_{\rm 3f, min})
\label{I3fsolution}
\end{equation}
with the limits

\begin{eqnarray}
\overline t_{\rm 3f, min} =& \max\left\lbrace \overline t_f \; , \;
{1 \over \nu_0} \left( {1 \over \gamma} - {1 \over \gamma_{\rm 1,f}} \right) 
\right\rbrace \cr
\overline t_{\rm 3f, max} =& \min\left\lbrace \overline t \; , \; 
\left( {1 \over \nu_0} \, \left[ {1 \over \gamma}
- {1 \over \gamma_{2, f}} \right] + \overline{\Delta t}_{\rm acc, f} \right) 
\right\rbrace \cr
\label{I3flimits}
\end{eqnarray}

\section{\label{sscphotons}Evaluation of the synchrotron photon density
for SSC}

For each incoming photon direction ($\pm$), the seed space- and time-dependent 
seed photon density $\overline n_{\rm ph}^{\pm} (\overline\epsilon_s \ox, 
\overline t_{\rm x, em})$ consists of contributions from all three branches
of the electron distribution (\ref{ne}), and may contain contributions from
both the forward and reverse shock. For any given branch of the electron
distribution, the calculation is analogous for the various cases to be
considered. Therefore, we give here only one representative case for
each branch.

The space- and time-dependent photon density distribution is evaluated as

\begin{equation}
\overline n_{\rm ph}^{\pm} (\overline\epsilon_s, \overline x, \overline t_{\rm x, em})
= {\sigma_T \over 48 \pi^2 \overline\epsilon_s^{1/2} \, m_e c^2}
\int\limits_{\overline x_{\rm s, min}^{\pm}}^{\overline x_{\rm s, max}^{\pm}} \,
{B(\overline x')^2 \over b(\overline x')^{3/2}} \, n_e \left( \sqrt{\overline\epsilon_s
\over b(\overline x')} \, , \, \overline x' \, , \, \overline t_x' \right) \,
\ln\left( {R \over \vert \ox - \oxp \vert } \right)
\; d \overline x'
\label{nphplusminus}
\end{equation}
where 

\begin{equation}
\overline t_x' = \overline t - {\vert \overline x' \, \mu \vert \over 
\overline\beta_{f,r} c} + {\overline x \over \Gamma^2 c} - 
{\vert \overline x' - \overline x \vert \over c}.
\label{txprime}
\end{equation}
Although $B(\overline x')$ and $b(\overline x')$ are constant throughout the
forward shock region and the reverse shock region, we need to take into account 
that generally $B_f \ne B_r$. The integration limits in Eq. \ref{nphplusminus} 
are set by the conditions

\begin{eqnarray}
\overline x' \ge& \overline x & {\rm (a^+)} \;\;\;\; {\rm or} \cr
\overline x' \le& \overline x & {\rm (a^-)} \cr
\overline t_x' >& 0 & {\rm (b)} \cr
\overline x' \le& \overline{\Delta r}_a & {\rm (c)} \cr
\label{xplusconditions}
\end{eqnarray}
and hence

\begin{eqnarray}
\overline x_{\rm s, min}^+ =& \overline x \cr
\overline x_{\rm s, max}^+ =& \min\left\lbrace \overline{\Delta r}_a \, , \, 
\left( \betaf \, {c \, \overline t + \overline x \, \left[ {\mu \over \Gamma^2} + 1 
\right] \over \overline\beta_f + 1} \right) \right\rbrace
\label{xpluslimits}
\end{eqnarray}

\begin{eqnarray}
\overline x_{\rm s, min}^- =& - \min\left\lbrace \overline{\Delta r}_b \, , \, 
\left( \betar \, {c \, \overline t + \overline x \, \left[ {\mu \over \Gamma^2} - 1 
\right] \over \overline\beta_r + 1} \right) \right\rbrace \cr
\overline x_{\rm x, max}^- =& \overline x
\label{xminuslimits}
\end{eqnarray}

Because of the absolute values involved in $\overline t_x$ (Eq. \ref{txprime}),
we need to evaluate the integral in Eq. \ref{nphplusminus} separately for the
cases $\overline x > 0$ (A) and $\overline x < 0$ (B). 

For case (A), the integral in Eq. \ref{nphplusminus} can be split up into 
6 contributions:

\begin{equation}
I_{s}^{A\pm} \equiv \left( I_{s1}^{A\pm} + I_{s2}^{A\pm} + I_{s3}^{A\pm} \right)
\label{ISApm}
\end{equation}
corresponding to the forward and backward traveling photons from the three 
terms in the expression for the electron density in Eq. \ref{ne}. Hence,

\begin{equation}
I_{s1}^{A+} = Q_{0,f} \, {B_f^2 \over b_f^{3/2}} \gamma^{-q} 
\int\limits_{\overline x_{\rm s, min}^+}^{\overline x_{\rm s, max}^+} 
H(\gamma_{\rm up} - \gamma) \, H(\gamma; \gamma_1, \gamma_c) \, \
\min\lbrace \overline t_x, \overline{\Delta t}_{\rm acc, f} \rbrace
\, \ln\left( {R \over \vert \ox - \oxp \vert } \right) \, d\overline x'
\label{IS1Aplus}
\end{equation}
with $\gamma = \sqrt{\overline\epsilon_s / b_f}$. 

In the region corresponding to the A+ contribution, we parameterize the
synchrotron emission time as 

\begin{equation}
\overline t_x' \equiv \alpha_{\rm A+} - \beta_{\rm A+} \, \overline x'
\label{alphabetaplusdef}
\end{equation}
with
\begin{eqnarray}
\alpha_{\rm A+} =& \overline t + {\overline x \over c} \left( {\mu \over \Gamma^2}
+ 1 \right) \cr
\beta_{\rm A+} =& {\overline\beta_f + 1 \over \overline\beta_f c} 
\end{eqnarray}

The effective limits of the integration (\ref{IS1Aplus}) are determined 
by the Heaviside functions as

\begin{eqnarray}
\overline x_{\rm s1, min}^{A+} =& \max\left\lbrace \overline x_{\rm s, min}^+ \, ,
\, {1 \over \beta_{A+}} \left( \alpha_{A+} - {1 \over \nuf \, \gamma} + 
\max \lbrace 0 \, , \, [\nuf \, \gamma_2]^{-1} - \dtaccf 
\rbrace \right) \right\rbrace \cr
\overline x_{\rm s1, max}^{A+} =& \overline x_{\rm s, max}^+ \cr
\label{S1Apluslimits}
\end{eqnarray}

To evaluate the integral $I_{s1}^{A+}$ with these limits, we define a 
critical $\overline x'_c$ for which $\overline t_x (\overline x'_c) = 
\overline {\Delta t}_{\rm acc, f}$:

\begin{equation}
\xcaplus \equiv {\alpha_{A+} - \overline{\Delta t}_{\rm acc, f} 
\over \beta_{A+}}
\label{xcAplus}
\end{equation}
with $\otx < \dtaccf$ if $\oxp > \xcaplus$. Furthermore, we write

Then, the integration yields:

$$
I_{s1}^{A+} = \gamma^{-q} \, Q_{0,f} \, {B_f^2 \over b_f^{3/2}} \, 
H(\gamma - \gamma_{\rm 1,f}) \times 
$$
\begin{equation}
\cases{ R \, f_{s1}^{A+} (y) \Biggr\vert_{y_{\rm min}}^{y_{\rm max}} 
& for $\xcaplus < \overline x_{\rm s1, min}^{A+} < \overline x_{\rm s1, max}^{A+}$ 
\cr\cr
R \, f_{s1}^{A+} (y) \Biggr\vert_{y_c}^{y_{\rm max}} 
+ \overline{\Delta t}_{\rm acc, f} \, R \, g (y)
\Biggr\vert_{y_{\rm min}}^{y_c} 
& for $\overline x_{\rm s1, min}^{A+} < \xcaplus < \overline x_{\rm s1, max}^{A+}$ 
\cr\cr
\overline{\Delta t}_{\rm acc, f} \, R \, g (y) 
\Biggr\vert_{y_{\rm min}}^{y_{\rm max}} 
& for $\overline x_{\rm s1, min}^{A+} < \overline x_{\rm s1, max}^{A+} < \xcaplus$ 
\cr
}
\label{IS1Aplusresult}
\end{equation}

with
\begin{eqnarray}
y_{\rm min} =& {\ox_{\rm s1, min}^{A+} - \ox \over R} \cr
y_{\rm max} =& {\ox_{\rm s1, max}^{A+} - \ox \over R} \cr
y_c =& {\xcaplus - \ox \over R} \cr
\label{yAoneplus}
\end{eqnarray}
and
\begin{eqnarray}
f_{s1}^{A+} (y) =& (\beta_{A+} \, \ox - \alpha_{A+}) \, (y \, \ln y - y) + {\beta_{A+} 
\, R \, y^2 \over 2} \, \left( \ln y - {1 \over 2} \right) \cr
g (y) =& y - y \, \ln y \cr
\label{fgAoneplus}
\end{eqnarray}

Analogous calculations (carefully accounting for the different implications
of the absolute values in Eq. \ref{alphabetaplusdef} and the different magnetic-field
values in the forward- and reverse-shock regions) yields the contributions 
A- and B$\pm$ from the first term in Eq. \ref{ne}.

The $A+$ ($\oxp > \ox > 0$) contribution of the second term in Eq. \ref{ne} 
is

\begin{equation}
I_{s2}^{A+} = Q_{0,f} {B_f^2 \over b_f^{3/2}} \, \gamma^{-(1+q)} \, H(\gamma - 
\gonef) \, \int\limits_{\overline x_{\rm s, min}^+}^{\overline x_{\rm s, max}^+}
{H(\gamma; \gc, \gup) \, \min\lbrace \otx , \dtaccf \rbrace \over \nuf \, \otx} 
\, \ln\left( {R \over \oxp - \ox} \right) \, d\oxp
\label{IS2Aplus}
\end{equation}

The effective limits of this integration are
\begin{eqnarray}
\overline x_{\rm s2, min}^{A+} =& \max\left\lbrace \overline x_{\rm s, min}^+ \, ,
\, {1 \over \beta_{A+}} \left( \alpha_{A+} - {1 \over \nuf} \left[ {1 \over \gamma} 
- {1 \over \gtwof} \right] - \dtaccf \right) 
\right\rbrace \cr
\overline x_{\rm s2, max}^{A+} =& \min\left\lbrace \overline x_{\rm s, max}^+ 
\, , \, {1 \over \beta_{A+}} \left( \alpha_{A+} - {c \over \nuf \, \gamma} \right) 
\right\rbrace \cr
\label{S2Apluslimits}
\end{eqnarray}
Using the parametrization for $\otx$ from Eq. \ref{alphabetaplusdef} and the 
critical value of $\oxp$ from Eq. \ref{xcAplus} for which $\otx = \dtaccf$,
the solution to $I_{S2}^{A+}$ is

$$
I_{S2}^{A+} = {Q_{0,f} \, B_f^2 \over \nuf \, b_f^{3/2}} \, \gamma^{-(1+q)}
\, H(\gamma - \gonef) \times
$$
\begin{equation}
\cases{
\dtaccf \, f_{s2}^{A+} (y) \Biggr\vert_{y_{\rm min}}^{y_{\rm max}} & 
for $\ox_{\rm s2, min}^{A+} < \ox_{\rm s2, max}^{A+} < \xcaplus$ \cr
\dtaccf \, f_{s2}^{A+} (y) \Biggr\vert_{y_{\rm min}}^{y_c} + R \, g(y)
\Biggr\vert_{y_c}^{y_{\rm max}} & 
for $\ox_{\rm s2, min}^{A+} < \xcaplus < \ox_{\rm s2, max}^{A+}$ \cr
R \, g(y) \Biggr\vert_{y_{\rm min}}^{y_{\rm max}}
& for $\xcaplus < \ox_{\rm s2, min}^{A+} < \ox_{\rm s2, max}^{A+}$}
\label{ISA2plussolution}
\end{equation}
with the definition of $y_{\rm min,max,c}$ analogous to Eq. \ref{yAoneplus}, 
and

$$
f_{s2}^{A+} (y) = {\ln y \, \ln (\alpha_{A+} - \beta_{A+} \, [ \ox -
R \, y] ) \over \beta_{A+}} 
$$
\begin{equation}
- {1 \over \beta_{A}} \, \left( \ln [ \alpha_{A+}
- \beta_{A+} \, \ox ] + \ln y - {\beta_{A+} \, R \, y \over \alpha_{A+} -
\beta_{A+} \, \ox} \, \Phi \left[ {\beta_{A+} \, R \, y \over \alpha_{A+}
- \beta_{A+} \, \ox} \, , \, 2 \, , 1 \right] \right) 
\label{fAs2plus}
\end{equation}
where
\begin{equation}
\Phi (z, 2, 1) = \sum\limits_{n=0}^{\infty} (n + 1)^{-2} \, z^n
\end{equation}
is the Lerch function. 

Again, analogous calculations yield the A- and B$\pm$ contributions of the
second term in Eq. \ref{ne}.

The $A+$ ($\oxp > \ox > 0$) contribution of the third term in Eq. \ref{ne} 
is

\begin{equation}
I_{s3}^{A+} = {Q_{0,f} \, B_f^2 \, \dtaccf \over b_f^{3/2}} \, \gonef^{-q} \,
\gamma^{-2} \, \int\limits_{\ox_{\rm s, min}^+}^{\ox_{\rm s, max}^+} 
H(\gup - \gamma) \, \, H(\gamma; \gmin , \gonef ) \, \ln\left( {R \over \oxp - \ox} 
\right) \, d\oxp
\label{IS3Aplus}
\end{equation}
The effective integration limits are given by

\begin{eqnarray}
\ox_{\rm s3, min}^{A+} =& \max\left\lbrace \ox_{\rm s, min}^+ \, , \, {1 \over
\beta_{A+}} \, \left( \alpha_{A+} - {1 \over \nuf} \, \left[ {1 \over \gamma} - 
{1 \over \gtwof}\right] - \dtaccf \right) \right\rbrace \cr
\ox_{\rm s3, max}^{A+} =& \min\left\lbrace \ox_{\rm s, max}^+ \, , \, {1 \over
\beta_{A+}} \, \left( \alpha_{A+} - {1 \over \nuf} \, \left[ {1 \over \gamma} 
- {1 \over \gonef} \right] \right) \right\rbrace \cr
\label{xs3pluslimits}
\end{eqnarray}
and the result of the integration is then

\begin{equation}
I_{s3}^{A+} = {Q_{0,f} \, B_f^2 \, \dtaccf \over b_f^{3/2}} \, \gonef^{-q} \,
\gamma^{-2} \, H(\gonef - \gamma) \, R \, \left( g[y_{\rm max}] - g[y_{\rm min} \right)
\label{IS3Aplusresult}
\end{equation}
with
\begin{eqnarray}
y_{\rm min} =& {\ox_{\rm s3, min}^{A+} - \ox \over R} \cr
y_{\rm max} =& {\ox_{\rm s3, max}^{A+} - \ox \over R} \cr
\end{eqnarray}

Analogous solutions are obtained for the A- and B$\pm$ contributions.

\end{document}